\documentclass[aps,prb,twocolumn,showpacs,floatfix,superscriptaddress]{revtex4-1}

\usepackage{graphicx}
\usepackage{latexsym}

\usepackage{subfigure}
\usepackage{hyperref}
\usepackage{epsfig}
\usepackage{float}
\usepackage{amsmath}
\usepackage{amssymb}
\usepackage{wasysym}
\usepackage{lipsum}
\usepackage{color}
\usepackage[toc,page]{appendix}
 
\usepackage[dvipsnames]{xcolor}

\usepackage{soul}

\newcommand{\bfq}{\mathbf{k}}

\newcommand{\beq}{\begin{equation}}
\newcommand{\eeq}{\end{equation}}

\newcommand{\rr}{\textbf{r}}
\newcommand{\ud}{\mathrm{d}}
\newcommand{\qq}{\mathbf{q}}

\newcommand{\tS}{{\tilde{S}}}

\begin{document}
\title{Sleuthing out exotic quantum spin liquidity in the pyrochlore magnet Ce$_2$Zr$_2$O$_7$}

\author{Anish Bhardwaj}
\affiliation{Department  of  Physics, Florida  State  University,  Tallahassee,  FL  32306,  USA}
\affiliation{National High Magnetic Field Laboratory,  Tallahassee,  FL  32310,  USA}

\author{Shu Zhang}
\affiliation{Department of Physics \& Astronomy, University of California, Los Angeles, CA, USA}

\author{Han Yan}
\affiliation{Department of Physics \& Astronomy, Rice University, Houston, TX 77005, USA}

\author{Roderich Moessner}
\affiliation{Max Planck Institute for Physics of Complex Systems, 01187 Dresden, Germany}

\author{Andriy H. Nevidomskyy}
\affiliation{Department of Physics \& Astronomy, Rice University, Houston, TX 77005, USA}

\author{Hitesh J. Changlani}
\affiliation{Department  of  Physics, Florida  State  University,  Tallahassee,  FL  32306,  USA}
\affiliation{National High Magnetic Field Laboratory,  Tallahassee,  FL  32310,  USA}
  
\date{\today}
\begin{abstract}
\end{abstract}

\maketitle

\textbf{
%\lipsum[2-4]
The search for quantum spin liquids  (QSL) -- topological magnets with fractionalized excitations -- has been a central theme in condensed matter and materials physics. While theories are no longer in short supply, 
tracking down materials  has turned out to be remarkably tricky, in large part because of the difficulty to diagnose  experimentally a state with only topological, rather than conventional, forms of order.  
Pyrochlore systems have proven particularly promising, hosting a classical Coulomb phase in the spin ices Dy/Ho$_2$Ti$_2$O$_7$\cite{Fennell2009,Morris2009}, with subsequent proposals of candidate QSLs in other pyrochlores. 
Connecting experiment with detailed theory exhibiting a robust QSL has remained a central challenge. 
Here, focusing on the strongly spin-orbit coupled effective $S=1/2$ 
pyrochlore Ce$_2$Zr$_2$O$_7$, we analyse 
recent thermodynamic and neutron scattering experiments, to identify 
a microscopic effective Hamiltonian through a combination of 
%finite temperature Lanczos, Monte Carlo, analytical and spin dynamics 
finite temperature Lanczos, Monte Carlo and analytical spin dynamics
calculations. Its parameter values suggest a previously unobserved exotic phase, a $\pi$-flux U(1) QSL. Intriguingly, the octupolar nature of the moments makes them less prone to be affected by crystal imperfections or magnetic impurities, while also hiding some otherwise characteristic signatures from neutrons, making this QSL arguably more stable than its more conventional counterparts.}

\begin{figure*}[tp]
\includegraphics[width=\linewidth]{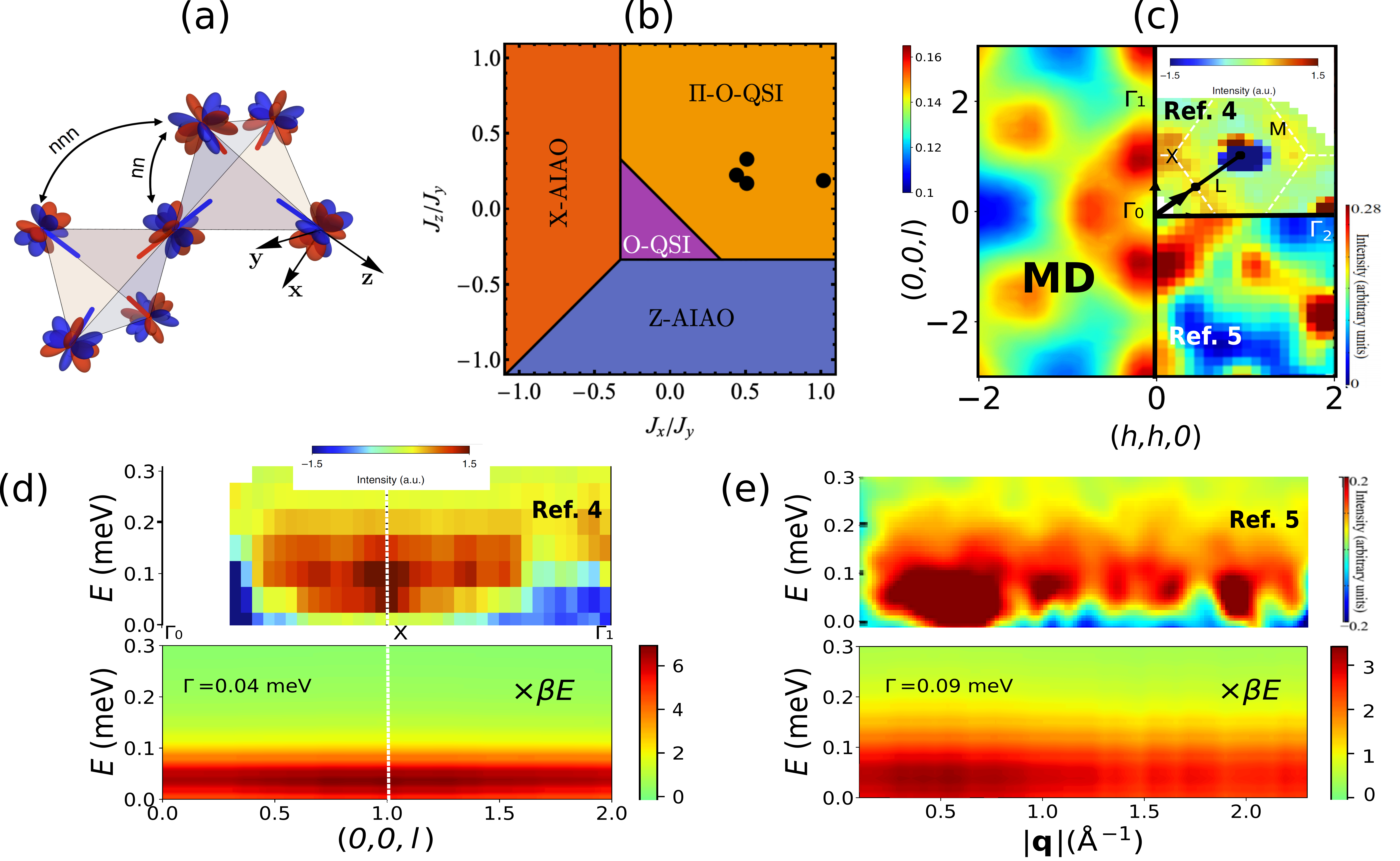}
\caption{
\textbf{Quantum spin liquid in a pyrochlore magnet and its experimental ramifications}.  \textbf{a} Depiction of a pyrochlore lattice with octupolar components forming 2-in-2-out ice states, and showing representative nearest neighbor (nn) and next nearest neighbor (nnn) bonds. \textbf{b}
Our model parameter sets in black dots, superposed on the mean
field phase diagram of Ref.~\onlinecite{Patri-octupolarQSI}.  %represented by the numbers 1,2,3 and 4 
\textbf{c} Dynamical structure factors at $T=0.06$ K integrated over the energy range 0.00 to 0.15 meV obtained from classical Molecular Dynamics (MD) for 8192 spins along with the corresponding comparison with Ref.~\onlinecite{CZO.Gao} ($T=0.035$ K) and Ref.~\onlinecite{CZO.Gaulin} ($T=0.06$ K, with an additional background subtraction). The plots reported in the published references were adapted for the purpose of comparison with our results. Special points in the Brillouin zone are also shown. \textbf{d,e} The dynamical structure factor as function of energy and momentum using the MD data (for 1024 sites) compared to experiment. Note that the quantum-classical correspondence dictates that the MD data be rescaled~\cite{ncnf.zhang} by the factor $ \beta E$ ($\beta = 1/k_B T$ where $k_B$ is the Boltzmann factor, and $E$ is the neutron energy transfer).
Panel \textbf{d} is for the momentum $(0,0,l)$ cross section ($\Gamma_0 \rightarrow X \rightarrow \Gamma_1$) and panel \textbf{e} is  the powder average. For panels \textbf{c,d,e} the numerically optimized Hamiltonian (parameter set no. 2 with both nearest neighbor and next nearest neighbor terms, see SM) was simulated. Additional Lorentzian convolution of width $\Gamma$ was applied to mimic the limited experimental resolution, based on values reported in Ref.~\onlinecite{CZO.Gao} and Ref.~\onlinecite{CZO.Gaulin}. The color scheme used for the theoretical calculation and the two experiments  differs, and in the absence of additional information only the relative variations should be compared.  At $(\pm 1, \pm 1,\pm 1)$ nuclear contributions have been effectively subtracted out in panel \textbf{c} using the high-temperature data, however this subtraction is not perfect, and residual intensity is seen at these positions. The same is true in panel (e) at $|\textbf{q}| \sim 1\,\text{\AA}^{-1}$ and $|\textbf{q}| \sim 2\,\text{\AA}^{-1}$, where the maxima of intensity originate from imperfect subtraction of nuclear Bragg peaks, which are absent from the theoretical calculations.
%\hjc{The large weight at $|\textbf{q}| \sim 1$~\AA and $|\textbf{q}| \sim 2$ \AA~is an artefact of experimental high temperature subtraction errors to eliminate nuclear Bragg peaks~\cite{CZO.Gaulin}, an issue our magnetic model does not address.} 
%The theoretical calculations do not have any nuclear Bragg contributions.
}

%shows the integrated dynamical structure factor from Molecular Dynamics simulations. Both (c) and (d) were obtained by using the dipole-octupole Hamiltonian with next nearest neighbor interaction term. Panels \textbf{e} and \textbf{f} were obtained from Ref~\onlinecite{CZO.Gao} and Ref~\onlinecite{CZO.Gaulin} respectively and these represent the  experimental dynamical structure factor (integrated over the energy range 0.05 to 0.15 mev). \an{Note that the nuclear Bragg peaks at $(\pm 1, \pm 1,\pm 1)$ have been subtracted in panels (e) and (f) using the high-temperature data, however this subtraction is not perfect, and the residual intensity is seen at these positions (especially visible as orange-red in panel f), whereas the theoretical calculations in panels (c) and (d) do not have any nuclear Bragg contributions.}}

\label{fig:fig1}
\end{figure*}

\iffalse
Pyrochlore family of materials such as exemplified by Ho$_2$Ti$_2$O$_7$ or Dy$_2$Ti$_2$O$_7$ famously provide realizations of classical spin ices~\cite{Ramirez_1999}, so named in analogy with the arrangement of molecules in the water ice~\cite{Bernal_Fowler_1933, Pauling1935}, with 
large classical degeneracy of the spin-ice manifold resulting in entropy-driven classical spin liquid. 
If one is able to make the rare-earth magnetic moments smaller (as in Tb, Ce, or Yb ions), quantum effects  become important, with the resultant  ``quantum spin ice" expected to form a quantum spin liquid with gapless excitation spectrum~\cite{hermele}. Despite several proposals over the past decade, no conclusive experimental realization of such QSL has been found.  
It is in this context that when lack of magnetic ordering and liquid-like structure of neutron scattering response were discovered in Ce$_2$Zr$_2$O$_7$ two years ago~\cite{CZO.Gao, CZO.Gaulin}, it was immediately proposed that this material epitomizes the long sought-after quantum spin ice. Nevertheless, pinning down what kind of QSL it is and deducing the parameters of the effective model that describe this quantum liquid is far from an easy task.
\fi

\begin{figure}[t!]
     \includegraphics[scale=0.371]{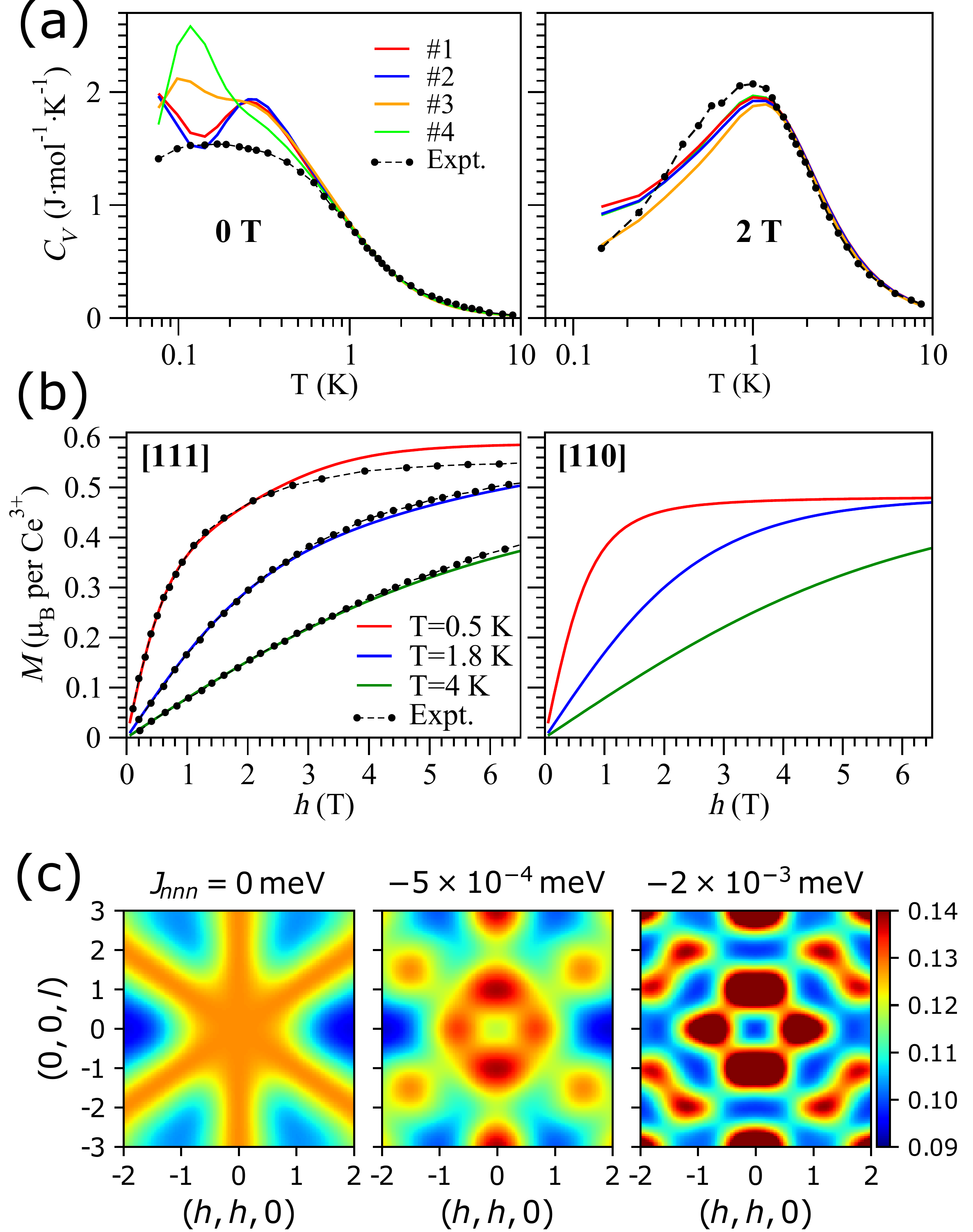}
\caption{\label{fig:fig2} \textbf{Fitting magnetization, specific heat and neutron scattering data to obtain the model Hamiltonian parameters.} Results of a two stage fitting process used to determine the optimal Hamiltonian consistent with all reported experiments. \textbf{a} Specific heat as a function of temperature obtained for four different model parameter sets, vs.\  experimental data for the case of zero field and applied field of $2$~T along the [111] direction. \textbf{b} Magnetization obtained from finite temperature Lanczos of the model Eq.~\eqref{eq:Hnn} on a 16-site cluster as functions of magnetic field applied along the [111] and [110] directions. For clarity, the theoretical data is shown for one parameter set only (set no. 2, see \textbf{Methods} and SM, also for fits with other parameter sets, of comparable quality) and only the experimental results for magnetization with field along [111] direction has been shown.
\textbf{c} The predicted zero-field neutron-scattering structure factor $S(\qq)$, Eq.~\eqref{eq:Sk}, computed with SCGA, for three different values of $J_{nnn}$, with the left panel ($J_{nnn}=0$) 
corresponding to the nearest-neighbor only model, Eq.~\eqref{eq:Hnn}. 
%Note the expected absence of the pinch-points in $S(\kk)$.
}
\end{figure}

The lack of magnetic ordering and liquid-like structure of neutron scattering response in Ce$_2$Zr$_2$O$_7$~\cite{CZO.Gao, CZO.Gaulin}, immediately led to its proposal as the long sought-after QSL known as quantum spin ice.
What makes the cerium-based Ce$_2$Zr$_2$O$_7$ pyrochlore distinct from the earlier studied Yb-based quantum spin ice candidate  Yb$_2$Ti$_2$O$_7$~\cite{YTO.Ross, YTO.Scheie, YTO.applegate} 
is the dipolar-octupolar nature~\cite{huang2014} %\cite{DO.GangChen} 
of the ground state doublet of Ce$^{3+}$ ion, shown schematically in Fig.~\ref{fig:fig1}(a). It is, to a very high accuracy, given by the $|J=5/2, m_J=\pm 3/2\rangle$ doublet,  %\rim{we're doubly using $J_z$ here, see Hamiltnian} 
where the quantization axis $z$ is chosen as the local $[111]$ axis of the cubic lattice~\cite{CZO.Gao}. The transverse components of the angular momentum 
%$J_x$ and $J_y$ 
thus have vanishing expectation values in the ground state, which in turn implies that they are invisible to the spin-flip scattering in the neutron scattering experiments because $\Delta m_J = 3$ excitations do not couple, to leading order, to the dipolar magnetic moment of neutrons. Using an effective pseudospin 1/2 representation of the ground state doublet~\cite{huang2014}, one of the three components (historically denoted $s^y$) represents the octupolar moment, and the other two components ($s^x$ and $s^z$) transform like the familiar dipole spinors.

The key to the unusual properties of Ce$_2$Zr$_2$O$_7$ are the  
effective interactions between spin components, both dipolar and octupolar, belonging to the nearest-neighbours Ce ions, which are different from the thoroughly studied dipolar spin ice.
%\hy{[(sugggested modification, HY) -- A physical consequence  of the unusual properties of Ce$_2$Zr$_2$O$_7$ is the allowed interaction ..., which is different from the thoroughly-studied dipolar case.]}.
A model Hamiltonian incorporating all symmetry-allowed spin-spin interactions on a tetrahedron is~\cite{huang2014}:
\beq
H_{nn} = \sum_{\langle ij\rangle} J_y s_i^y s_j^y + [J_x s_i^x s_j^x + J_z s_i^z s_j^z + J_{xz}(s_i^xs_j^z + s_i^z s_j^x)],
\label{eq:Hnn}
\eeq
with $s_i$ again expressed in the local frame (relative to the local 
% $\langle 111\rangle$ 
$[111]$
direction on a given Ce site).

By fitting the experimental magnetization and specific heat to (quantum) finite temperature Lanczos method (FTLM) calculations~\cite{FTLM.Prelovsek}
~(see Fig.~\ref{fig:fig2} and Methods for more details), we have determined the parameters of the model Hamiltonian in Eq.~\eqref{eq:Hnn}. A crucial result from the modeling perspective is that we have identified the $J_y$ interaction|which acts between the octupolar components|to be the largest term ($J_y \approx 0.1$~meV), with two interactions ($J_x$ and $J_z$)  playing a subleading role, and a vanishing $J_{xz}$.
We have obtained several sets of parameter values within the fitting error bars, shown by the black dots in Fig.~\ref{fig:fig1}(b), clustered around $J_y = 0.08\pm 0.01$~meV, $J_x = 0.05 \pm 0.02$~meV, $J_z = 0.02 \pm 0.01$~meV.
%The fitted parameter values are shown  by the black dots in Fig.~\ref{fig:fig1}(b), corresponding to the sets of parameters we obtained, up to numerical fitting errors \hy{[this sentence reads a bit repetitive]}. 
We show additional cost function analyses for a wide range of $J_x, J_y$ and $J_z$ in the SM, that illustrate constraints on our fits given the current availability of experimental data. %In particular, we find that while individual values of $J_x$ and $J_z$ have relatively large error bars, their sum is well constrained: $J_x + J_z = 0.066 \pm 0.010$~meV.

Note that since the octupolar $s^y$ moments do not couple to neutron spins in the leading order, it is crucial to perform fits to the magnetization and specific heat as described above; attempts to %instead
fit solely the dynamical structure factors measured in inelastic neutron scattering (INS) are less reliable.
%\rim{suggest dropping: and can lead to erroneous conclusions.} 
% What we have below is factually incorrect - HJC
% Andriy starts
%Instead, we use the results obtained from fitting the specific heat and magnetization (see Fig.~\ref{fig:fig2}) to compute the predicted INS spectra and compare them with the experiments. This comparison, both for the energy integrated and energy-resolved data, is shown in Fig.~\ref{fig:fig1} \textbf{c,d,e}. 
% Andriy ends
% HJC rewrite using Andriy's words where possible
We do employ the INS data, however, at the second stage of our fitting process. After determining $H_{nn}$, we introduce an additional weak next-nearest neighbor (nnn) coupling $J_{nnn}$, which likely originates from the magnetic dipole-dipole interaction between Ce ions.
% We compare the Self Consistent Gaussian Approximation (SCGA) prediction of the spin structure factor with INS data to determine its optimal value (see Fig.~\ref{fig:fig2}). We find $J_{nnn} \sim 0.005 J_y \approx 0.5\; \mu$eV, which likely originates from the dipole-dipole interaction between Ce ions. 
Its optimal value $J_{nnn} \sim 0.005 J_y \approx 0.5\; \mu$eV was determined by comparing the Self Consistent Gaussian Approximation (SCGA) prediction of the spin structure factor with INS data (see Fig.~\ref{fig:fig2}(c)). 
%}\sz{Just a tiny rearrangement of the sentence.}

% HJC rewrite using Andriy's sentences where possible
Armed with this complete Hamiltonian (nn + nnn), we compute the energy-integrated and energy-resolved momentum dependent neutron spin structure factor using classical Molecular Dynamics (MD)~\cite{Conlon_Chalker,ncnf.zhang, LL.Samarakoon}. Representative comparisons with previous experiments on Ce$_2$Zr$_2$O$_7$~\cite{CZO.Gao,CZO.Gaulin} are shown in Fig.~\ref{fig:fig1}(c-e). Panel (c) shows the characteristic ring-like structure centered around $\Gamma_0 = (0,0,0)$ point, with pronounced maxima at $\mathbf{q}=(0,0,1)$ and $(\frac{1}{\sqrt{2}},\frac{1}{\sqrt{2}},0)$, consistent with experimental observations, where we use the standard Miller indices $(h,k,l)$ to denote the direction in reciprocal space.
%\sz{You mean at $\mathbf{q} = (0,0,\pm1)$? 
% I find this very confusing because Miller indices usually denote crystalline planes.
Note that the high intensity points at $\mathbf{q}=(1,1,\pm1)$ and $(2,2,-2)$ seen in the experimental panels in Fig.~\ref{fig:fig1}(c) result from an imperfect subtraction of the nuclear Bragg peaks, which are absent in our magnetic model. 
Using the quantum-classical correspondence to rescale the MD data~\cite{ncnf.zhang}, in panel (d) we show the results for a one-dimensional cross section in momentum space ($\Gamma_0 \rightarrow X \rightarrow \Gamma_1$). The increased intensity at the $X=(001)$ point at low energies is broadly consistent with experimental findings. Panel (e) shows our results for the powder averaged case, compared with the experimental data in Ref.~\onlinecite{CZO.Gaulin}, suggesting an overall agreement of the energy scales over the entire Brillouin zone (the $X$ point corresponds to $|\mathbf{q}| \sim 0.6\,\text{\AA}^{-1}$ where the intensity is highest). 
Note that the large weight seen in the experimental data in Fig.~\ref{fig:fig1}(e) at $|\textbf{q}| \sim 1\,\text{\AA}^{-1}$ and $|\textbf{q}| \sim 2\,\text{\AA}^{-1}$ is an artefact of an imperfect subtraction of the high temperature data to eliminate the nuclear Bragg peaks~\cite{CZO.Gaulin}, an issue our magnetic model does not address. 

\begin{figure}[t!]
\includegraphics[width=\columnwidth]{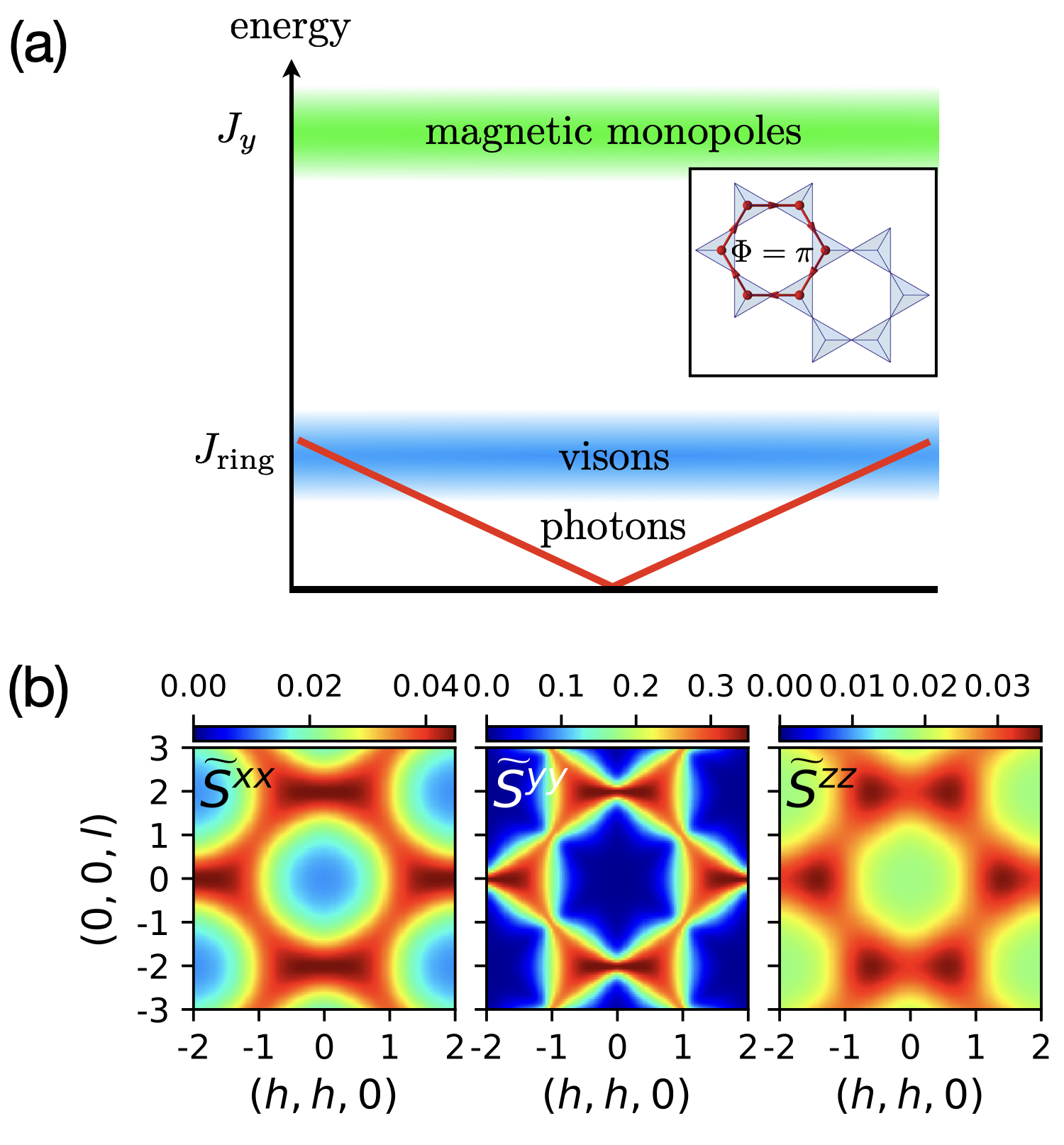}
\caption{\textbf{Properties of the octupolar quantum spin-ice state}.
\textbf{a} Schematic depiction of the energy spectrum of $\pi-$OQSI state, with the green band denoting the spin-flip excitations that create a pair of magnetic monopoles (spinons). The blue band represents the dispersive, gapped vison excitations,
which are analogous to electric charges. 
The red line is the hallmark of gauge-neutral photon excitations of the quantum spin-ice, except here the octupolar (rather than dipolar) character renders the photons `invisible' to neutrons.
\textbf{Inset,} Representation of the gauge flux $\Phi$ through the hexagon formed by 6 corner-sharing tetrahedra on the pyrochlore lattice. \textbf{b} The diagonal pseudospin correlation functions
$\tilde{S}^{xx}$, $\tilde{S}^{yy}$ and $\tilde{S}^{zz}$  as in
$\tilde{S}^{\alpha\beta}(\qq) = \int \ud \rr\, e^{-i\qq\cdot \rr} \langle s^\alpha(\rr) s^\beta(0) \rangle$.
Here $x,y,z$ are the local axes (relative to the local $\hat{\mathbf{z}}=
% \langle 1 1 1\rangle
[111]$
direction). The pinch-point patterns in the $\tilde{S}^{yy}$ channel are invisible to neutrons because of the octupolar nature of $s^y$.  
} 
\label{fig:OQSI}
\end{figure}

\begin{figure}[ht]
\begin{center}
\vskip 0.3 in
\includegraphics[width=\linewidth]{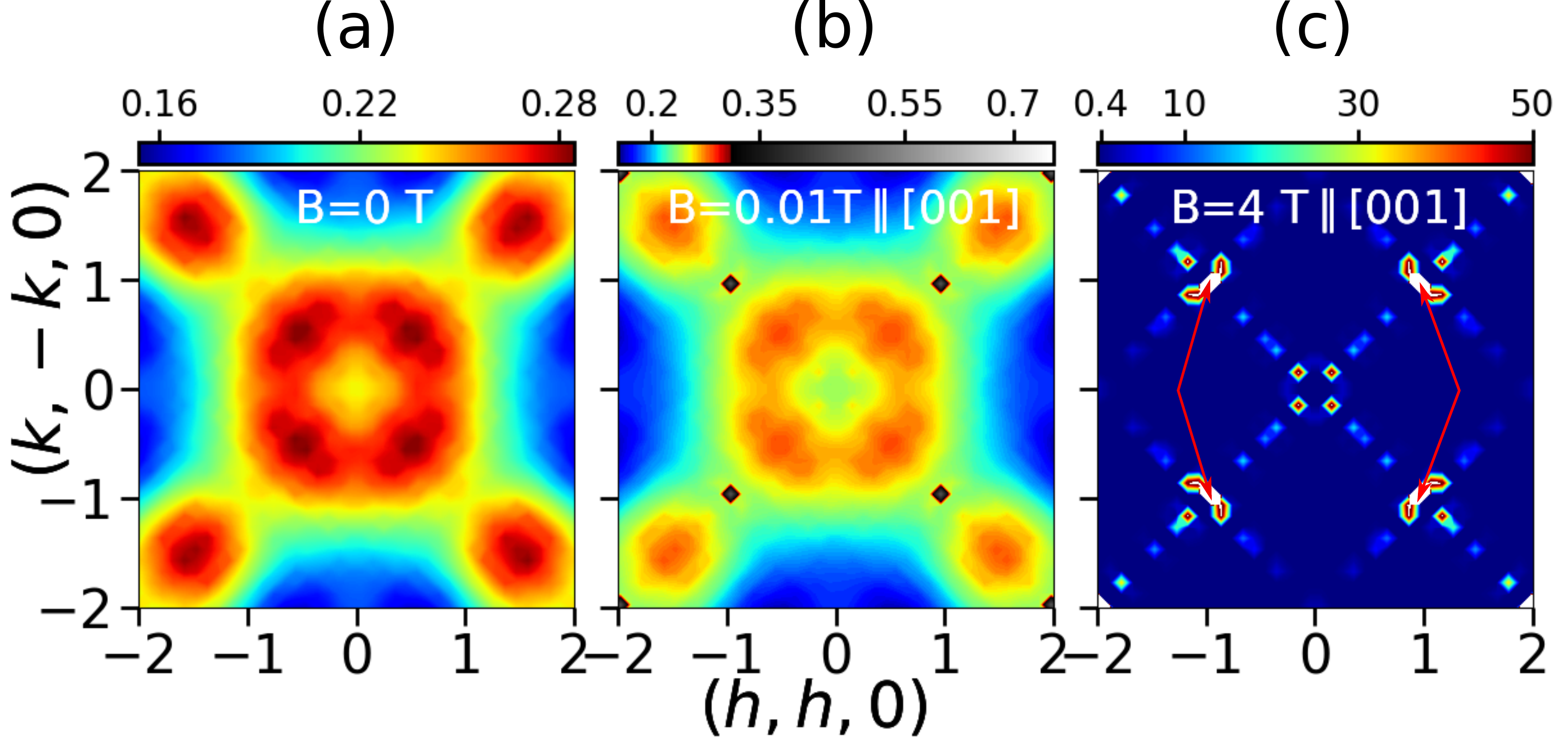}
\caption{\textbf{Predictions for static spin structure factor as expected to be measured in INS.} The predicted (energy-integrated) neutron-scattering spin structure factor (in arbitrary units) 
in zero field and an applied magnetic field along 
[001] 
$\mu_0 H=0.01$~T and  $\mu_0 H=4$~T, computed using MC calculations.
The Bragg peaks at $4$ T, which have intensity value $\sim$ 3400, appear in the white regions indicated with arrows.  
}
\label{fig:4}
\end{center}
\end{figure}

%\an{\textbf{A.N. will explain the concept of octupolar spin-ice here.}}

{\bf Discussion:} 
The fact that 
%the inelastic neutron scattering signal is 
multiple experimental features are accurately reproduced by a model of  dipolar-octupolar interactions between the Ce spin components
% magnetic moments
in Eq.~\eqref{eq:Hnn} 
(plus a small $J_{nnn}$ term)
poses the question about what phase corresponds to its ground state. 
Indeed, the fact that the coupling $J_y$ between the nearest neighbor octupolar moments is antiferromagnetic and by far the largest suggests that the leading behaviour is for the corresponding moments to form a (classical) 2-in/2-out spin-ice manifold. The presence of non-zero $J_x$ and $J_z$ interactions then adds quantum effects; generically, this opens the possibility of obtaining a quantum spin ice phase. 

%\st{Indeed, the previous analysis of the model in Eq.~\eqref{eq:Hnn}, using  mean-field analytics and 16-site exact diagonalization numerics in Ref.~
%\onlinecite{Patri-octupolarQSI}, presented a phase diagram for the model Hamiltonian, with the candidate parameters we determined all falling comfortably into a $\pi$-flux octupolar quantum spin-ice ($\pi-$OQSI) phase, as depicted by black dots  in Fig.~\ref{fig:fig2}(b).} 
%\hy{[This sentence is very long. I suggest breaking it into shorter pieces. -- HY]}
The phase diagram for the model Hamiltonian in Eq.~\eqref{eq:Hnn} 
was studied in  Ref.~\onlinecite{Patri-octupolarQSI,Benton2020PhysRevB}, 
using a combination of analytical and  mean-field analysis as well as  exact diagonalization.
In our modeling of the magnetization and specific heat, we have determined four candidate sets of fitting parameters (subject to the errorbars in fitting), depicted by black dots  in Fig.~\ref{fig:fig1}(b). All four fall deep into the parameter regime of the $\pi$-flux octupolar quantum spin-ice ($\pi-$OQSI) phase according to Ref.~\onlinecite{Patri-octupolarQSI}.
%, as depicted by black dots  in Fig.~\ref{fig:fig2}(b). }
In this phase, the emergent gauge field $a$  takes a non-trivial ground state configuration that hosts
a flux $\Phi=\pi$ through each hexagon\cite{Lee2012PhysRevB,GangChen2017PhysRevB,Benton2018PhysRevLett},  
\beq
(\nabla \times a)_{\hexagon} \equiv \sum_{i\in\hexagon} a_{\rr_i,\rr_{i+1}} = \pi,
\label{eq:pi-flux}
\eeq
as shown schematically in the inset of  Fig.~\ref{fig:OQSI}(a).
%\rim{what does the following sentence mean?} 
This fact follows from the effective U(1) quantum field theory, described in the Supplementary Materials, with the flux-dependent contribution in the form
\beq
H_\text{flux} = J_\text{ring} \sum_{\hexagon} \cos(\nabla \times a)_{\hexagon},
\label{eq:ring-exchange}
\eeq 
where $J_\text{ring} \sim (J_x + J_z)^3/(64 J_y^2)$ is positive and thus favors flux $\Phi = \pi$ in each hexagon, resulting in the $\pi-$OQSI phase. 

Like ``conventional'' quantum spin ice, 
 %\hy{[Suggested revision to make the paragraph more structured and also refer to the figures more accurately:] 
 $\pi-$OQSI has gapless photons and two types of gapped excitations (magnetic and electric charges), in close analogy to Maxwell electrodynamics. 
 Their approximate energy scales are illustrated in Fig.~\ref{fig:OQSI}(a).
 The first type of gapped excitations are spinons
 created in pairs by flipping the octupolar moment on a single site, costing energy around $\mathcal{O}(J_y)$.
 %\st{, creates a pair of oppositely charged spinons on neighbouring tetrahedra, illustrated schematically in Fig.~\ref{fig:OQSI}.} 
 These are analogues of the
 magnetic monopoles in electrodynamics, and observable in the specific heat as a characteristic Schottky peak at energy $\sim J_y \approx 1$~K, as our FTLM calculations corroborate in Fig.~\ref{fig:fig2}(a). 
The second type of gapped excitations corresponds to the so-called visons 
 (analogues of electric charges),  
%\st{(a.k.a. magnetic charges)} 
%\hy{[For the gauge theory language we use, visons are analog of electric charges. They are refereed to as magnetic charges only in the dual language.]}, 
which are sources of the gauge flux violating the condition in Eq.~(\ref{eq:pi-flux}). 
%This flux enters the Hamiltonian in the form 
Note the energy scale for exciting visons at $J_\text{ring} \sim 0.03$~K is very low. One therefore expects to find thermally excited visons  even at the base temperature of the experiment, so that their gap, if not closed by the quantum dynamics,  will not be separately resolved. 
Rather, visons will strongly interact and mix with  
the emergent gapless  photons, named in analogy to the photons familiar from Maxwell electrodynamics, 
%\st{as illustrated schematically in} 
due to the overlap of their energy scales [cf.
Fig.~\ref{fig:OQSI}(a)].

The energy and temperature scales for observing the photons would correspondingly be very low. More importantly, they would not directly couple to neutrons because of the octupolar nature of the $\pi-$OQSI. To illustrate this point, we have computed the pseudospin correlation functions $\tilde{S}^{\alpha\beta}(\qq) = \int \ud \rr\, e^{-i\qq\cdot \rr} \langle s^\alpha(\rr) s^\beta(0) \rangle$ in Figure~\ref{fig:OQSI}(b) by SCGA (see \textbf{Methods} for details). The largest components of this matrix are the diagonal ones $\tS^{xx}$, $\tS^{yy}$ and $\tS^{zz}$. It is the octupolar $\tS^{yy}(\qq)$ component in the middle panel of Figure~\ref{fig:OQSI}(b) that displays the pinch-points characteristic of the spin-ice \cite{moessner98-PRB58,Isakov2004PhysRevLett}, and the photons' gapless dispersion will emanate from the $\qq$ location of those pinch-points. 
Crucially however, the $s^y$ components of spin do not couple to the magnetic field or to neutron moment, as explained in \textbf{Methods} (see  Eq.~\eqref{eq:g-factors}), meaning that the aforementioned pinch-points will not feature in the experiment. Instead, the neutron-scattering structure factor 
\beq
S({\bf{q}}) \sim \sum_{\mu\nu} \left(\delta_{\mu\nu} - \frac{q_\mu q_\nu}{q^2}\right)\! \int \ud r\, e^{-i{\bf{q}}\cdot \rr}\langle m^\mu(
\rr) m^\nu(0) \rangle \!
\label{eq:Sk}
\eeq
is expressed in terms of true magnetic moments \mbox{$m^\mu = \sum_\lambda g^{\mu\lambda} s^\lambda$} that contain the  $g$-factors, whose  $g^{\mu y}$ components are all zero (see SM).  As a consequence, $\tilde{S}^{yy}$ drops out of the neutron structure factor, computed in Fig.~\ref{fig:fig1}(c) and Fig.~\ref{fig:fig2}(c) using    Monte Carlo and SCGA respectively.
The main contributors to the 
neutron scattering structure factor are the $\tilde{S}^{xx}$ and $\tilde{S}^{zz}$ channels 
convoluted with the neutron-coupling form factor in Eq.~\eqref{eq:Sk}.
As a result, instead of the pinch-points, a sixfold, 
three-rod-crossing-like
pattern is observed. 
Such rod pattern is expected for a pyrochlore lattice with nearest-neighbour interactions only~\cite{CastelnovoPhysRevB2019} and is associated with the dispersion of spinons (magnetic monopoles) in the context of quantum spin ice \cite{natphys.Sibille2018,KatoPhysRevLett,CastelnovoPhysRevB2019}.

Upon inclusion of (weak) nnn interactions $J_{nnn}$, the pattern of 
crossing rods deforms into
characteristic ring-like structure that appears in $S(\qq)$, as shown in the three panels of Fig.~\ref{fig:fig2}(c). Phenomenologically fixing the value $J_{nnn}\sim 0.5\;\mu$eV  matches very well with the experimental INS observations [see Fig.~\ref{fig:fig1}(c)]. 
Thence, while it is tempting to associate the
intensity variation along the rods in Fig.~\ref{fig:fig2}(c) as the disappearance of pinch point intensity centered at $(1,1,1)$ (and equivalent points), 
which has been predicted to be the quintessential feature of the dispersive, quantum photon modes in dipolar
quantum spin ice\cite{Benton2012PhysRevB},
our analysis rather suggests that these are the consequence of the  small  nnn interactions between Ce ions that modulate the INS intensity along the $(1,1,1)$ direction.
As for the emergent photons, while they are indeed expected to be present in the $\pi$-OQSI phase, as shown in Fig.~\ref{fig:OQSI}(a), their octupolar nature turns out to render them 
much less visible to neutrons,
and they can only be detected via weaker, higher-order coupling to neutrons at large momentum transfer \cite{natphys.sibille,Lovesey2020PhysRevB},
or indirectly, for instance through their contribution to the low-temperature specific heat (at $T\lesssim J_\text{ring}/k_B \approx 0.03$~K).

An interesting question is how the octupolar quantum spin-ice state responds to the application of an external magnetic field \cite{Benton2020PhysRevB}. While the octupolar $s^y$ pseudospin components do not couple linearly to the field and will remain in the 2-in/2-out configuration, the $s^z$ components will cant along the field direction (see Eq.~\eqref{eq:g-factors}, where $g_x\approx 0$). In order to elucidate the experimental consequences, we compute the in-field spin-structure factor $S(\qq)$ within classical Monte Carlo calculations on our model, shown in Fig.~\ref{fig:4}. As a function of increasing field along $[001]$ direction, the ring-like structure in $S(\mathbf{q})$ quickly weakens (Fig.~\ref{fig:4}b), until eventually disappearing and giving way to sharp Bragg peaks in high fields (Fig.~\ref{fig:4}c). These predictions are to be compared with future INS data in an applied magnetic field.

The fact that the magnetic octupolar degrees of freedom do not couple in the leading order to neutron spin or to the external magnetic field, makes the octupolar spin liquid difficult to detect. Its elusive nature may however prove to be a blessing in disguise, as a reduced coupling to magnetic defects and associated stray magnetic fields -- which are known to destabilize the more conventional dipolar spin liquids -- are similarly suppressed. Indeed, chemical disorder on magnetic sites is believed to be the leading reason for the failure to observe the quantum spin liquid behaviour in, for instance, the herberthsmithite  kagome compounds despite their high crystallographic quality~\cite{norman-herbertsmithite}. The fact that neither magnetic order {\it nor} spin glassiness is seen in cerium pyrochlores Ce$_2$Zr$_2$O$_7$ and Ce$_2$Sn$_2$O$_7$~\cite{natphys.sibille,prl.sibille} may be taken as an additional, albeit indirect, evidence of the robustness of the underlying octupolar spin liquid.  While the possibility of such a quantum spin liquid has been entertained in seminal theoretical studies  before~\cite{huang2014,Patri-octupolarQSI}, our present work firmly identifies Ce$_2$Zr$_2$O$_7$ as a very promising host for the $\pi$-flux octupolar quantum spin ice phase. 
The present study also underscores the importance of carefully fitting multiple experiments, including specific heat and magnetization, in addition to the INS spectra, to determine the effective model Hamiltonian, which otherwise may be plagued with uncertainties affecting the searches and identification of  QSLs~\cite{Maksimov2020,Laurell2020}.

\bibliography{ref}

\begin{thebibliography}{33}
\bibitem [{foo(note)}]{footnote-H3}%
  \BibitemOpen
  \href@noop {} {} (\bibinfo {year} {footnote}),\ \bibinfo {note} {the
  octupolar moments can in principle couple to the third power of the magnetic
  field $H^3$, however this effect is negligible for the experimentally
  accessible fields.}\BibitemShut {Stop}%
\end{thebibliography}
%\bibliographystyle{nature}

%\section*{Methods}
\vspace{4mm}
\noindent
\textbf{Methods}
\vspace{1mm}
\newline\noindent
{\small
\textbf{Fitting model Hamiltonian parameters:}
In addition to the nearest neighbor (nn) Hamiltonian in Eq.~\eqref{eq:Hnn} we have considered the next nearest neighbor (nnn) interaction, which in the local basis is given by,  
\begin{eqnarray}
H_{nnn}=\sum_{\langle\langle ij \rangle \rangle } J_{nnn}  \left[s_i^x~s_i^y~s_i^z\right]\left[\begin{array}{ccc}
 {g_x^2} & 0 & {g_x}{g_z}\\
0 & 0 &0\\
 {g_x}{g_z} & 0 &{g_z^2}
\end{array}\right]\left[\begin{array}{c}
s_j^x \\ 
s_j^y \\
s_j^z 
\end{array}\right].\nonumber\\
\end{eqnarray}

For the case of an applied external magnetic field, the Zeeman term must be also be accounted for. This term involves the coupling of magnetic field to effective spin 1/2 degrees of freedom, which are not the usual and familiar dipoles, and are instead dipolar-octupolar doublets. 

A key observation is that the octupolar magnetic moments 
%($s^y$ in our notation)
do not couple, to linear order, to the external magnetic field.~\cite{footnote-H3} 
The only coupling of the external magnetic field is to the dipolar degrees of freedom. It can be shown that only the component of the field $H$ along the local $z$-axis (i.e. local [111] direction) couples to the dipole  moment~\cite{huang2014}, as follows:
\beq
\label{eq:g-factors}
H_Z = -\sum_i (\mathbf{h}\cdot \mathbf{z}_i)(g_z s_i^z + g_x s_i^x),
\eeq
where {$|\mathbf{h}|=\mu_B \mu_0 H$ is the effective magnetic field strength, $\mu_B$ is the Bohr magneton. Note that when projected onto the $J=5/2$ multiplet, one expects the Land\'e g-factor $g_z=2.57$ and $g_x = 0$. However, if an admixture of higher spin-orbit multiplet ($J=7/2$) is present in the ground state doublet, one generically expects a non-zero value of $g_x$ (see Supplementary Materials), which we allow for in our modeling. 

As explained in the main text, the Hamiltonian parameters in Eq.~\eqref{eq:Hnn} were determined using both zero and applied magnetic field data. For the specific heat and magnetization, we have performed quantum 
FTLM calculations on a 16-site cluster. Details of the general technique can be found in Ref.~\onlinecite{FTLM.Prelovsek} and previous application to some pyrochlore systems can be found in Ref.~\onlinecite{Changlani.YTO}. 
Convergence checks of the method have been discussed at length in the SM. 
Results of the fitting process are shown in Fig.~\ref{fig:fig2}. 
Fig.~\ref{fig:fig2}(a) shows the temperature dependence of the specific heat and suggests that fitting it is somewhat challenging, both in zero and applied field.
Our four distinct parameter sets are generally able to describe specific heat very well at higher temperature $T\gtrsim 1$~K, however the lower-temperature behavior is trickier and none of the parameter sets used satisfactorily fits the data especially in the absence of applied magnetic field. We attribute this to the finite-size effects in our FTLM calculations, which become more pronounced at lower temperatures. The magnetization as a function of applied field strength, shown in Fig.~\ref{fig:fig2}(b), appears to be less prone to finite size effects and is fitted reasonably well, especially at low fields. 
The discrepancy at high fields is not entirely unexpected, previous experimental reports have suggested a changing $g$ factor past a $\mu_0 H=4$ T 
field~\cite{CZO.Gao}, an effect not built into our model.

Despite the limitations to do with the finite-system size in the numerics, and the finite energy and momentum resolution in INS experiment, all parameter sets obtained are in agreement 
with the antiferromagnetic $J_y$ which we find is large compared to the other two interactions ($J_x$ and $J_z$) in Eq.~\eqref{eq:Hnn}, as depicted in Fig.~\ref{fig:fig1}(b). To further build confidence in our results, we have performed a brute force scan of $J_x$ and $J_z$ for representative fixed values of $J_y$ and constructed a contour map of an appropriately defined cost function (see SM, more subtleties with the fitting and additional competitive parameter sets are also discussed).
Additional future experiments could potentially further constrain the values of these coupling constants. 

The second step of our parameter fitting involved the determination of $J_{nnn}$ which we found to be small relative to $J_y$. Despite its smallness, it is responsible for significant reorganization of intensity in the Brillouin zone. Fig.~\ref{fig:fig2}(c) shows the static structure factor in the $(h,h,l)$ plane computed with SCGA (the details of which will be explained shortly) for representative values of $J_{nnn}$. Performing a brute force line search, we determined the optimal value $J_{nnn} = - 5 \times 10^{-4}$ meV working in steps of $10^{-4}$ meV. 

The following values of the parameters (which we refer to as set no. 2, see SM) were used in subsequent calculations (all values are in meV): $J_x=0.0385, J_y=0.088, J_z=0.020, J_{xz} = 0, J_{nnn} = -0.0005 $, with the $g$-factors $g_x=-0.2324$, $g_z=2.35$. The other parameters sets are quoted in full in the SM.

\vspace{2mm}
\noindent
\textbf{Details of the Monte Carlo and Molecular (spin) Dynamics calculations:} 
The dynamical structure factor is computed by integrating the classical Landau-Lifshitz equations of motion
\begin{equation}
        \frac{d}{dt} \mathbf{S}_i = -\mathbf{S}_i\times \frac{\partial H}{\partial \mathbf{S}_i},
\label{eq:torque}
\end{equation}
which describes the precession of the spin in the local exchange field. We carry out all our calculations by transforming our Hamiltonian to the global basis, and have used the label ${\bf S_i}$ to represent spins in this basis (see SM for more details on transformations between local and global bases).

Following the protocol adopted in previous work~\cite{Conlon_Chalker, ncnf.zhang}, the initial configuration (IC) of spins is drawn by a Monte Carlo (MC) run from the Boltzmann distribution $\exp(-\beta H)$ at temperature $T \!=\! 0.06~K$. Then, for each starting configuration the spins are deterministically evolved according to Eq.~(\ref{eq:torque}) with the fourth order Runge-Kutta method. This procedure, referred to as molecular dynamics (MD), is repeated for many independent IC (their total number being $N_{IC}$) 
and the result is averaged,
\begin{eqnarray}
\langle S^{\mu}_i(t) S^{\nu}_j(0) \rangle =  \frac{\sum_{\text{IC from MC}} S^{\mu}_i(t) S^{\nu}_j(0) \big|_\text{IC}}{N_{IC}}.
\label{eq:mc-md}
\end{eqnarray}
We perform a Fourier transform in spatial and time coordinates to get the desired dynamical structure factor. We work with $N\!=\!16L^3$ sites, where $L^3$ is the number of cubic unit cells, the results in Fig.~\ref{fig:fig1}(c) are for $L\!=\!8$ ($N=8192$) and $L=4$ ($N=1024$) for Fig.~\ref{fig:fig1}(d,e). $N_{IC} \approx 10^4$ was used, and each IC was evolved for $500~\mathrm{meV}^{-1}$ in steps of $\delta t \!=\! 0.02$ meV$^{-1}$.

To obtain an estimate of the quantum dynamical structure factor, we used a classical-quantum correspondence, which translates to a simple rescaling of the classical MD data by $\beta E$. More details and justification can be found in Ref.~\onlinecite{ncnf.zhang}.

\vspace{1mm}
\noindent
\textbf{Convolution with Lorentzian function to mimic limitations of experimental resolution:} 
Fig.~\ref{fig:fig1}(d,e)) shows significant broadening along the energy axis, 
an effect not captured to the same extent by the raw (rescaled) MD data. Since the interaction energy scales in the material are small, a fairer comparison between experiment and theory is achieved by modeling the instrument's energy resolution. Using $\Gamma$ values in the ballpark suggested by Refs.~\onlinecite{CZO.Gao, CZO.Gaulin}, we convolved our data with a Lorentzian factor,
\begin{equation}
    S_\text{exp}(\qq,E) = \frac{1}{\pi}\int S(\qq,E')_{MD} \frac{\Gamma}{\Gamma^2 + (E' - E)^2} dE'
\end{equation}
The integral was approximated by a sum over discrete $E'$ points in steps of 0.01 meV.

\vspace{1mm}
\noindent
\textbf{Details of the SCGA calculations:} 
The Self-Consistent Gaussian Approximation 
is an analytical method that treats
the spin in the Large-N limit.
Our calculation follows closely 
Ref.~\onlinecite{Isakov2004PhysRevLett}.
In this study, 
we first treat $s^{x,y,z}$ as independent, freely fluctuating degrees of freedom.
The Hamiltonian in momentum space is written as
\begin{equation}
\mathcal{E}_\text{Large-N} = 
 \frac{1}{2}\mathbf{S}\mathcal{H}_\text{Large-N} \mathbf{S}^T ,
\end{equation}
where $\mathbf{S} = (s_1^x, s_2^x, s_3^x, s_4^x,\dots,  s_3^z, s_4^z).$
The interaction matrix ${H}_\text{Large-N}$
is the Fourier transformed interaction matrix that includes the nearest and next nearest neighbor interactions.

We then introduce a Lagrangian multiplier
with coefficient $\mu$
to the partition function
to get
\begin{equation} 
    \mathcal{Z} =   \exp\left( -\frac{1}{2} {\int_\text{BZ}  \text{d}\bfq \text{d}\mathbf{S} \  \mathbf{S}  \left[\beta  \mathcal{H}_\text{Large-N}+ \mu
	 \mathcal{I}\right]\mathbf{S} }  \right)   
\end{equation}
in order to impose an additional constraint
of averaged spin-norm being one, or
\begin{equation}
  \langle  \mathbf{s}_1^2  +  \mathbf{s}_2^2 +   \mathbf{s}_3^2 +   \mathbf{s}_4^2 \rangle  = 1 .
\end{equation}
For a given temperature $k_B T= 1/\beta$, the value of $\mu$ is fixed by this constraint via relation 
\begin{equation}
\int_\text{BZ}   \text{d} \bfq \sum_{i=1}^{12}\frac{1}{\lambda_i(\bfq)+\mu} = \langle \mathbf{s}_1^2 + \mathbf{s}_2^2 +\mathbf{s}_3^2 +\mathbf{s}_4^2   \rangle = 1   ,
\end{equation}
where $\lambda_i(\bfq),\ i=1,2,\dots, 12$
are the twelve eigenvalues of $\beta\mathcal{H}_\text{Large-N}$.
With $\mu$ fixed,
the partition function 
is completely determined for a free theory
of $\mathbf{S}$,
and all correlation functions can be computed from $\left[\beta  \mathcal{H}_\text{Large-N}+ \mu
\mathcal{I}\right]^{-1}$.\\

\noindent
\textbf{Data Availability}\newline\noindent
{\small
The data analyzed in the present study is available from the first author (A.B.) upon reasonable request.
}\\
  
\noindent
\textbf{Acknowledgements}\newline\noindent
{\small
We acknowledge useful discussions with J. Gaudet.
H.Y. and A.H.N. acknowledge the support of the National Science Foundation Division of Materials Research under the Award DMR-1917511.
Research at Rice University was also supported by the Robert A. Welch Foundation Grant No.~C-1818.  A.B. and H.J.C. thank Florida State University and the National High Magnetic Field Laboratory for support. The National High Magnetic Field Laboratory is supported by the National Science Foundation through NSF/DMR-1644779 and the state of Florida. H.J.C. was also supported by NSF CAREER grant DMR-2046570.  S.Z. was supported by NSF under Grant No. DMR-1742928. 
This work was partly supported by the Deutsche Forschungsgemeinschaft under grants SFB 1143 (project-id 247310070) and the cluster of excellence ct.qmat (EXC 2147, project-id 390858490)
A.H.N. and R.M. acknowledge the hospitality of the Kavli Institute for Theoretical Physics (supported by the NSF Grant No. PHY-1748958), where this work was initiated. A.H.N. thanks the Aspen Center for Physics, supported by National Science Foundation grant PHY-1607611, where a portion of this work was performed. 
We thank the Research Computing Cluster (RCC) and Planck cluster at Florida State University for computing resources.}\\
 
\noindent
\textbf{Author contributions}\newline\noindent
{\small
A.H.N. and R.M. conceived the theoretical ideas behind the project and planned the research. 
A.B., S.Z. and H.J.C. conceived and carried out the analysis of the experimental data and extraction of the effective Hamiltonian, as detailed in the Methods section and in the Supplementary Materials. A.B. and H.J.C. performed the finite temperature Lanczos, classical Monte Carlo and molecular (Landau-Lifshitz) spin dynamics calculations. S.Z. and H.Y. performed the self-consistent Gaussian calculations. 
All authors contributed to discussion and interpretation of the results. A.B., S.Z., H.Y. and H.J.C. prepared the figures. A.H.N. and R.M. wrote the manuscript with contributions from all authors.}\\

%\section*{Competing interests}
\noindent
\textbf{Competing interests}\newline\noindent
{\small
The authors declare no competing interests.
}\\

%\vspace{4mm}

%\section*{Additional information}
\noindent
\textbf{Additional information}\newline\noindent
{\small
}

\appendix
\clearpage
\newpage
%\beginsupplement
\setcounter{equation}{0}
\setcounter{figure}{0}
\setcounter{table}{0}
\setcounter{page}{1}
\makeatletter
\renewcommand{\bibnumfmt}[1]{[#1]}
\renewcommand{\citenumfont}[1]{#1}

%%%%%%%%%% Merge with supplemental materials %%%%%%%%%%
\begin{widetext}
\begin{center}
	\Large{Supplementary Materials for ``Sleuthing out exotic quantum spin liquidity in the pyrochlore magnet Ce$_2$Zr$_2$O$_7$"}
\end{center}
\end{widetext}

\section{Effective Hamiltonian: $J$- and $g$-matrices}
Adopting a notation similar to that used in Refs.~\onlinecite{huang2014,Patri-octupolarQSI}, the most general nearest-neighbor (nn) Hamiltonian that describes the dipole-octupole system in terms of effective spin-1/2 degrees of freedom (defined in a local basis) is given by,
\begin{eqnarray}
\label{eq:Hnn_supp}
H_{nn}&=&\sum_{\langle ij \rangle} J_x s_i^x s_j^x+J_ys_i^ys_j^y+J_zs_i^zs_j^z +J_{xz}(s_i^xs_j^z+s_i^zs_j^x)   \nonumber \\
&&-\underset{i}{\sum}\left(\hat{z}_i\cdot {\bf{h}}
\right)(g_xs_i^x+g_zs_i^z) 
\end{eqnarray}
In this convention $s_i^x$ and $s_i^z$ refer to dipolar degrees of freedom and $s_i^y$ refers to the octupolar degree of freedom on site $i$. $\langle ij \rangle$ refers to nn bonds. $J_x, J_y, J_z$ and $J_{xz}$ are interaction parameters and $g_x$ and $g_z$ denote coupling strengths of the dipolar degrees to the local $z$ component of ${\bf h}=\mu_B \mu_0 {\bf H}$, where ${\bf H}$ is the applied magnetic field and $\mu_B$ is the Bohr magneton.  Note that the last term differs from the usual Zeeman coupling of dipoles to an applied magnetic field.

Using sublattice labels $0,1,2,3$, for the four sublattices of the pyrochlore lattice, the local coordinate system at each site is given by,
\begin{eqnarray}
\hat{z}_0&=&\frac{1}{\sqrt{3}}(1,1,1),~~\hat{y}_0=\frac{1}{\sqrt{2}}(0,1,-1) \nonumber \\
\hat{z}_1&=&\frac{1}{\sqrt{3}}(1,-1,-1),~~\hat{y}_1=\frac{1}{\sqrt{2}}(-1,0,-1) \nonumber \\
\hat{z}_2&=&\frac{1}{\sqrt{3}}(-1,1,-1),~~\hat{y}_2=\frac{1}{\sqrt{2}}(-1,-1,0) \nonumber \\
\hat{z}_3&=&\frac{1}{\sqrt{3}}(-1,-1,1),~~\hat{y}_3=\frac{1}{\sqrt{2}}(-1,1,0) 
\end{eqnarray}  
Using a right handed coordinate system, the local $\hat{x}$ axis is given by  $\hat{x}_i = \hat{y}_i \times \hat{z}_i$.  

In order to compute observables, it is convenient to transform spins in the local basis to the global frame by using the relation,
\begin{eqnarray}
\left[\begin{array}{c}
s_i^x \\
s_i^y \\
s_i^z 
\end{array}\right]=R_i\left[\begin{array}{ccc}
S_i^x \\
S_i^y \\
S_i^z 
\end{array}\right].
\end{eqnarray}
where $R_i$ represents a rotation matrix on site $i$, which depends only on the sublattice it belongs to. $R_i$ are 
given by the expressions,
\begin{subequations}
\begin{eqnarray}
&&R_0=\left[\begin{array}{ccc}
\sqrt{\frac{2}{3}}& -\frac{1}{\sqrt{6}}& -\frac{1}{\sqrt{6}}\\
0 & \frac{1}{\sqrt{2}} & -\frac{1}{\sqrt{2}}\\
\frac{1}{\sqrt{3}} & \frac{1}{\sqrt{3}} &\frac{1}{\sqrt{3}}
\end{array}\right], \\%\nonumber \\
&&R_1=\left[\begin{array}{ccc}
-\frac{1}{\sqrt{6}} & -\sqrt{\frac{2}{3}} & \frac{1}{\sqrt{6}}\\
-\frac{1}{\sqrt{2}} & 0 & -\frac{1}{\sqrt{2}}\\
\frac{1}{\sqrt{3}} & -\frac{1}{\sqrt{3}} &-\frac{1}{\sqrt{3}}
\end{array}\right], \\% \nonumber\\
&&R_2=\left[\begin{array}{ccc}
\frac{1}{\sqrt{6}}& -\frac{1}{\sqrt{6}} & -\sqrt{\frac{2}{3}}\\
-\frac{1}{\sqrt{2}} & -\frac{1}{\sqrt{2}} & 0\\
-\frac{1}{\sqrt{3}} & \frac{1}{\sqrt{3}} &-\frac{1}{\sqrt{3}}
\end{array}\right], \\
&&R_3=\left[\begin{array}{ccc}
\frac{1}{\sqrt{6}}& \frac{1}{\sqrt{6}} & \sqrt{\frac{2}{3}}\\
-\frac{1}{\sqrt{2}} & \frac{1}{\sqrt{2}} & 0\\
-\frac{1}{\sqrt{3}} & -\frac{1}{\sqrt{3}} &\frac{1}{\sqrt{3}}
\end{array}\right]. 
\end{eqnarray} 
\end{subequations}

Using these expressions, the Hamiltonian in Eq.~\eqref{eq:Hnn_supp} in global basis acquires the form
\begin{eqnarray}
H_{nn}=\underset{\langle ij\rangle}{\sum}J_{ij}^{\mu\nu}S_i^{\mu}S_j^{\nu}-h^{\mu}\underset{i}{\sum}g_i^{\mu\nu}S_i^{\nu},
\label{hamil-global-app}
\end{eqnarray}
where $\mu,\nu$ refer to global Cartesian components $x,y,z$, and 
\begin{eqnarray}
J_{01}=\left[\begin{array}{ccc}
J_1 & J_2 &-J_1\\
J_3 & -J_1 &J_4\\
-J_4 & -J_1 &-J_3
\end{array}\right] && 
J_{02}=\left[\begin{array}{ccc}
-J_1 & J_1 & J_2\\
J_4 & J_3 & -J_1\\
-J_3 & -J_4 & -J_1
\end{array}\right] \nonumber\\
J_{03}=\left[\begin{array}{ccc}
-J_1 & -J_1 & -J_2\\
J_4 & -J_3 & J_1\\
-J_3 & J_4 & J_1
\end{array}\right] &&
J_{12}=\left[\begin{array}{ccc}
-J_3 & -J_4 & -J_1\\
J_1 & -J_1 & -J_2\\
-J_4 & -J_3 & J_1
\end{array}\right] \nonumber\\
J_{13}=\left[\begin{array}{ccc}
-J_3 & J_4  & J_1\\
J_1 & J_1  & J_2\\
-J_4 & J_3 & -J_1
\end{array}\right] &&
J_{23}=\left[\begin{array}{ccc}
-J_4 & J_3 & -J_1\\
-J_3 & J_4 & J_1\\
J_1 & J_1 & J_2
\end{array}\right]
\label{jmatrices-dip-oct-app}
\end{eqnarray}
%Here, 
where $J_1$, $J_2$, $J_3$ and $J_4$ are given by, %defined below
\begin{subequations}
\begin{eqnarray}
J_1&=&\frac{1}{6}\left(-2J_x+\sqrt{2}J_{xz}+2J_z\right), \\
J_2&=&\frac{1}{3}\left(-2J_x-2\sqrt{2}J_{xz}-J_z\right), \\
J_3&=&\frac{1}{6}\left(J_x-2\sqrt{2}J_{xz}-3J_y+2J_z\right),\\
J_4&=&\frac{1}{6}\left(-J_x+2\sqrt{2}J_{xz}-3J_y-2J_z\right).
\end{eqnarray} 
\end{subequations}
In a similar way, the $g$-matrices are given by
\begin{subequations}
\begin{eqnarray}
g_0&=&\left[\begin{array}{ccc}
g_+ & g_- & g_-\\
g_+ & g_- & g_-\\
g_+ & g_- & g_-
\end{array}\right],\\
g_1&=&\left[\begin{array}{ccc}
 g_- & -g_+ & -g_-\\
-g_- &  g_+ &  g_-\\
-g_- &  g_+ &  g_-
\end{array}\right], \\
g_2&=&\left[\begin{array}{ccc}
 g_- & -g_- &  g_+\\
-g_-&   g_- & -g_+\\
 g_- & -g_- &  g_+
\end{array}\right],\\
g_3&=&\left[\begin{array}{ccc}
g_- & g_- & -g_+\\
g_- & g_- & -g_+\\
-g_-&-g_- &  g_+
\end{array}\right].
\label{gmatrices}
\end{eqnarray}
\end{subequations}
where, $g_+=\frac{1}{3}(\sqrt{2}g_{x}+g_z)$ and $g_-=\frac{1}{3}(g_{z}-\frac{g_x}{\sqrt{2}})$.
These expressions for the $J$ and $g$ matrices differ from their more familiar dipolar counterpart~\cite{YTO.Ross}. 

%As explained in the main text, we also include the dipolar interaction for next-nearest neighbors (nnn).
The dipolar interaction between sites $i,\ j$ located at positions $\bf{r_i}$ and $\bf{r_j}$, respectively is 
\begin{equation}
    \frac{\vec{m}_i\cdot\vec{m}_j - 3(\vec{m}_i\cdot \hat{r}_{ij}) (\vec{m}_j\cdot \hat{r}_{ij})}{r_{ij}^3}.
\label{eq:dipolar}
\end{equation}
where $r_{ij}=|\bf{r_i - r_j}|$ is the distance between sites, $\hat{r_{ij}}$ is the unit vector along $\bf{r_i - r_j}$. 
$\vec{m}_i$ is the effective magnetic moment,
\begin{equation}
\vec{m}_i = \hat{z}_i (g_z s_i^z + g_x s_i^x).
\end{equation}
%The next nearest neighbor interaction takes 
Truncating Eq.~\eqref{eq:dipolar} to include only next-nearest neighbor terms, (the nearest neighbor pieces can be incorporated into $J_x$ and $J_z$), 
the Hamiltonian takes the form,
\begin{eqnarray}
H_{nnn}=\sum_{\langle\langle ij \rangle \rangle } J_{nnn}  \left[s_i^x~s_i^y~s_i^z\right]\left[\begin{array}{ccc}
 {g_x^2} & 0 & {g_x}{g_z}\\
0 & 0 &0\\
 {g_x}{g_z} & 0 &{g_z^2}
\end{array}\right]\left[\begin{array}{c}
s_j^x \\ 
s_j^y \\
s_j^z 
\end{array}\right].\nonumber\\
\label{eq:Hnnn_supp}
\end{eqnarray}
where $\langle\langle ij\rangle\rangle$ refers to next nearest neighbors (nnn) on the pyrochlore lattice and $J_{nnn}$ is the strength of the effective interactions.
%
%an adjustable parameter. As explained in the main text, we have determined the %optimal value of $J_{nnn}$ by fitting the static spin structure factor from %Self consistent Gaussian approximation (SCGA) calculations of the full %Hamiltonian (nn+nnn terms) to the experimental results available in the %literature~\cite{CZO.Gao, CZO.Gaulin}.

In the global basis, ~\eqref{eq:Hnnn_supp} takes the form
\begin{eqnarray}
H_{nnn}=\underset{\langle\langle ij \rangle \rangle}{\sum}J_{n_{ij}}^{\mu\nu}S_i^{\mu}S_j^{\nu}.
\end{eqnarray}
where $J_{n_{ij}}$ have the form,
\begin{subequations}
\begin{eqnarray}
J_{n_{01}}&=&\left[\begin{array}{ccc}
-J_{n_1} & -J_{n_3} &  J_{n_1}\\
 J_{n_2} &  J_{n_1} & -J_{n_2}\\
 J_{n_2} &  J_{n_1} & -J_{n_2}
\end{array}\right] \\
J_{n_{02}}&=&\left[\begin{array}{ccc}
 J_{n_1} & -J_{n_1} & -J_{n_3}\\
-J_{n_2} &  J_{n_2} &  J_{n_1}\\
-J_{n_2} &  J_{n_2} &  J_{n_1}
\end{array}\right] \\
J_{n_{03}}&=&\left[\begin{array}{ccc}
 J_{n_1} &  J_{n_1} &  J_{n_3}\\
-J_{n_2} & -J_{n_2} & -J_{n_1}\\
-J_{n_2} & -J_{n_2} & -J_{n_1}
\end{array}\right] \\
J_{n_{12}}&=&\left[\begin{array}{ccc}
-J_{n_2} &  J_{n_2} &  J_{n_1}\\
-J_{n_1} &  J_{n_1} &  J_{n_3}\\
 J_{n_2} & -J_{n_2} & -J_{n_1}
\end{array}\right] \\
J_{n_{13}}&=&\left[\begin{array}{ccc}
-J_{n_2} & -J_{n_2}  & -J_{n_1}\\
-J_{n_1} & -J_{n_1}  & -J_{n_3}\\
 J_{n_2} &  J_{n_2}  &  J_{n_1}
\end{array}\right] \\
J_{n_{23}}&=&\left[\begin{array}{ccc}
 J_{n_2} &  J_{n_2} &  J_{n_1}\\
-J_{n_2} & -J_{n_2} & -J_{n_1}\\
-J_{n_1} & -J_{n_1} & -J_{n_3}
\end{array}\right]
\label{jmatrices-dip-oct-app}
\end{eqnarray}
\end{subequations}
where $J_{n_1}$, $J_{n_2}$ and $J_{n_3}$ have been defined as,
%
%[The equations above are not labeled correctly. Are the following ones right?  -- Han]
\begin{subequations}
\begin{eqnarray}
J_{n_1}&=&\frac{J_{nnn}}{6}\left(2g_x^2-\sqrt{2}g_xg_z-2g_z^2 \right) ,\\
J_{n_2}&=&\frac{J_{nnn}}{6}\left(g_x^2-2\sqrt{2}g_xg_z+2g_z^2 \right) ,\\
J_{n_3}&=&\frac{J_{nnn}}{3}\left(2g_x^2+2\sqrt{2}g_xg_z+g_z^2 \right) .
\end{eqnarray}
\end{subequations}

\begin{figure}
\begin{center}
\includegraphics[width=\linewidth]{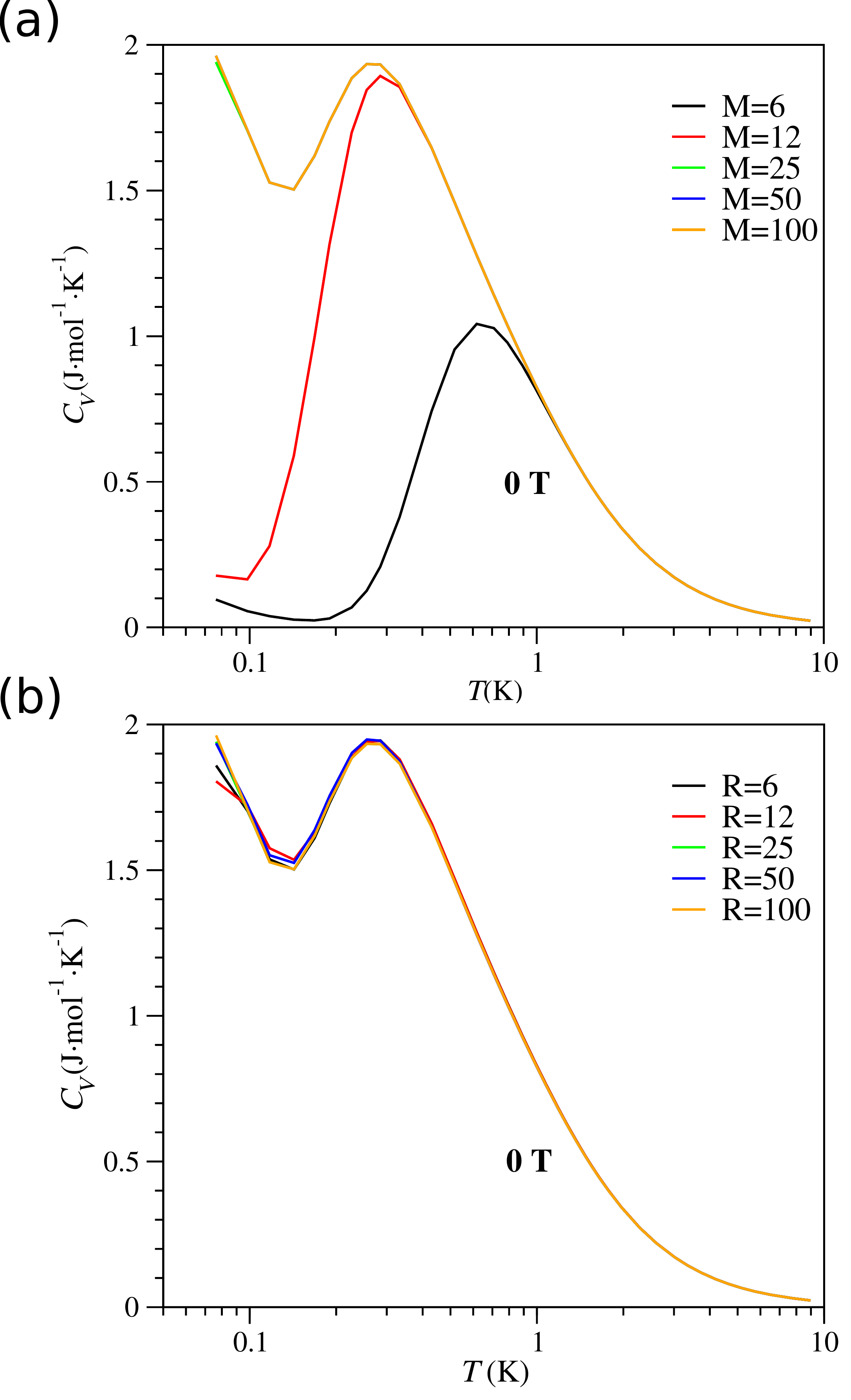}
\caption{Panel (a) shows the variation of the specific heat profile at zero field with $M$ for $R=100$ in the finite temperature Lanczos algorithm. Panel (b) shows the variation of the specific heat profile with $R$ for $M=50$. The Hamiltonian parameters for both panels correspond to set no. 2 (see Table~\ref{tab:H_params}).}
\end{center}
\label{fig:convergence}
\end{figure}

\section{Finite Temperature Lanczos Method}

\begin{figure*}
\begin{center}
\includegraphics[width=\linewidth]{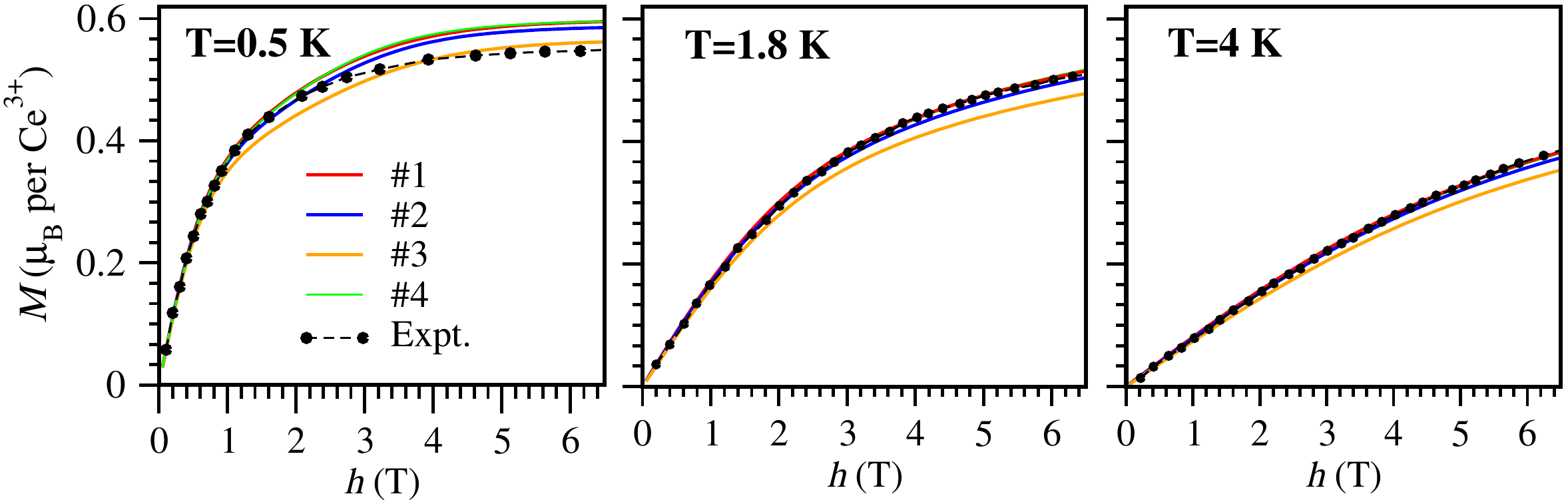}
\caption{Magnetization ($M$) vs field strength ($h$) for the [111] direction at temperatures of 0.5 K, 1.8 K and 4.0 K. The solid black circles connected by dashed lines represent the experimental data (extracted from ~\onlinecite{CZO.Gao}), the red, blue, orange and green solid lines represent the results from FTLM by using parameter set 1, 2, 3, and 4 respectively (see Table~\ref{tab:H_params}). The green and red curves visually overlap each other.}
\label{fig:M_params}
\end{center}
\end{figure*}

\begin{figure*}[htpb]
\begin{center}
 \includegraphics[width=\linewidth]{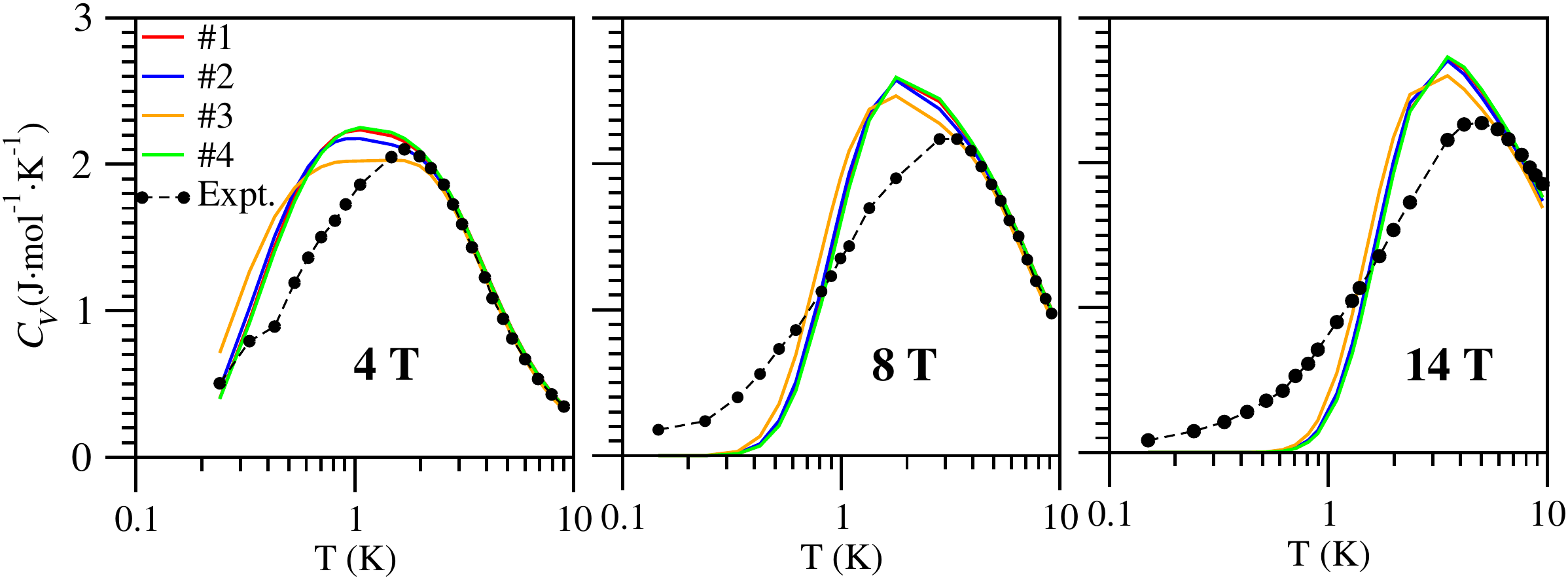}
\caption{Specific heat vs. temperature for four Hamiltonian parameter sets (shown in Table~\ref{tab:H_params}) as compared to experiment for field strengths of 4 T, 8 T and 14 T applied along the [111] direction. The experimental data was extracted from Ref.~\onlinecite{CZO.Gao}.}
\label{fig:C_params}
\end{center}
\end{figure*}

In the finite temperature Lanczos method (abbreviated as FTLM, see Ref.~\onlinecite{FTLM.Prelovsek} for details) the expectation value of any operator (A) is evaluated %commuting with the Hamiltonian 
using the expressions,
\begin{subequations}
\begin{eqnarray}
\label{eq:ftlm1}
\langle A\rangle&=&\frac{N_{st}}{ZR}\overset{R}{\underset{r=1}{\sum}}\overset{M}{\underset{j=0}{\sum}} e^{-\beta\epsilon_j^r}\langle r|\psi^r_j\rangle\langle\psi_j^r|A|r\rangle,\\
Z&=&\frac{N_{st}}{R}\overset{R}{\underset{r=1}{\sum}}\overset{M}{\underset{j=0}{\sum}}e^{-\beta\epsilon_j^r}|\langle r|\psi^r_j\rangle|^2 .
\end{eqnarray}
\end{subequations}
where $N_{st}$ is the dimension of the entire Hilbert space, $Z$ is the partition function and $\beta=1/k_B T$ is the inverse temperature. $|r\rangle$ is the initial random state, $R$ denotes the number of such starting states, and $M+1$ is the dimension of the Krylov space spanned by the vectors  $|r\rangle$, $H|r\rangle$, $H^2|r\rangle$,...,$H^{M}|r\rangle$.
$|\psi^r_j\rangle$ and $\epsilon^r_j$ represent (respectively) the $j^{th} $Ritz eigenvector and eigenvalue obtained by diagonalizing the Hamiltonian in the Krylov space. 
%spanned by the vectors  $|r\rangle$, $H|r\rangle$, %$H^2|r\rangle$,...,$H^{M-1}|r\rangle$.
%$R$ and $M$ denote the number of different random starting vectors $|r\rangle$ and the
%dimension of the Hamiltonian matrix in the Krylov space respectively. 

To evaluate the expectation value of the observables accurately the convergence with respect to both $R$ and $M$ were checked (see Fig.~\ref{fig:convergence} for a representative example and see previous work in Ref.~\onlinecite{Changlani.YTO}). The specific heat was evaluated using Eq.~\eqref{eq:ftlm1} to compute $\langle H \rangle$ and $\langle H^2\rangle$ and using the expression, %substituting them in the expression for specific heat given below
\begin{eqnarray}
C_v=\frac{1}{k_B T^2}\left(\langle H^2\rangle-\langle H\rangle ^2\right).
\end{eqnarray} 
Similarly, the magnetization was evaluated by calculating the free energy 
\begin{equation}
    F = -k_BT \ln Z
\end{equation}
and then taking its derivative with respect to the field strength.

\begin{table*}
\begin{center}
\begin{tabular}{|c|c|c|c|c|c|c|c|c|c|c|}
\hline
Set &$J_1$ & $J_2$ & $J_3$ & $J_4$ & $g_{x}$ & $g_{z}$ &$J_x$ & $J_y$ & $J_z$ & $J_{xz}$ \tabularnewline
\hline
parameter set 1 &-0.009 & -0.035  & -0.031  & -0.056 & 0 & 2.401 & 0.044 & 0.087 & 0.015 & 0 \tabularnewline
\hline
%parameter set 2 &-0.004 & -0.040  & -0.033  & -0.053 & -0.2324 & 2.35 & 0.039 & 0.086 & 0.021 & 0.007 \tabularnewline
%\hline
parameter set 2 &-0.006 & -0.032  & -0.030  & -0.057 & -0.2324 & 2.35  & $0.0385$ & $0.088$ & $0.020$ & $0$  \tabularnewline
\hline
%$parameter~set~3$ & 0.038 & -0.023  & -0.003  & -0.065 & 2.0 & 1.174 & -0.015 & 0.068  & 0.10 & %0\tabularnewline
parameter set 3 &-0.004 & -0.036  & -0.024  & -0.056 & 0.574 & 2.196 & 0.041 & 0.081  & 0.027  & 0  \tabularnewline
\hline
parameter set 4 &-0.018 & -0.05  & -0.018  & -0.049 & 0.0 & 2.4 & 0.069 & 0.068  & 0.013  & 0 \tabularnewline
\hline
\end{tabular}
\caption{Parameter sets studied in the paper. All $J$ values are in meV units, and all $g$ values are dimensionless}
\label{tab:H_params}
\end{center}
\end{table*}

\section{Parameter Fitting}
We devise a cost function involving the weighted errors between the experimental data (taken from Ref.~\onlinecite{CZO.Gao}) and the numerically computed observables for a given Hamiltonian parameter set. We use the specific heat which is known for different values of temperature and magnetic field strength $h$ and the magnetization along the [111] direction, and define,
%We construct the following cost function involving specific heat which is known for different values of temperature and magnetic field strength $h$
%\begin{eqnarray}
%f=\sum_j\sqrt{\frac{\underset{i=1}{\overset{N}{\sum}}(C_v^e(T_i,h_j)-C_v^s(T_i,h_j))^2}{N}}
%\end{eqnarray}
\begin{eqnarray}
f&=&\alpha_c \sum_j\sqrt{\frac{\underset{i=1}{\overset{N_c}{\sum}}(C_v^e(T_i,h_j)-C_v^s(T_i,h_j))^2}{N_c}}+\nonumber\\
&& \alpha_m \sum_j\sqrt{\frac{\underset{i=1}{\overset{N_m}{\sum}}(M^e(T_j,h_i)-M^s(T_j,h_i))^2}{N_m}}
\label{eq:cost_supp}
\end{eqnarray}
where $C_v^e, C_v^s$ are the specific heat and $M^e,M^s$ are the magnetization along the [111] direction from experiment (taken from Ref.~\onlinecite{CZO.Gao}) and FTLM simulations respectively. $N_c$ and $N_m$ are the number of data points for the specific heat and magnetization respectively.
$\alpha_c, \alpha_m$ are weight factors. For example, in situations where only specific heat fitting was of interest $\alpha_c=1$, $\alpha_m=0$. When discussing our various strategies, we specify what regimes of the experimental data were retained for the fitting procedure or for computing the cost function. 

It should be emphasized that minimizing the cost function in Eq.~\eqref{eq:cost_supp} without constraining the domain of each parameter can yield unphysical values. Additionally, given that the phase space spanned by the six parameters entering Eq.~\eqref{eq:Hnn_supp} is large, it can be time consuming to yield a meaningful solution. It is imperative that the optimization be performed by imposing meaningful bounds on each parameter.

These parameter bounds were found by noting that the specific 
heat curves for Ce$_2$Zr$_2$O$_7$ show no sharp anomaly at low 
temperature and the zero field specific heat has a Schottky 
like bump only for scales less than 0.3 K. This sets an 
approximate upper bound for the interaction strength at 0.10 
meV. If the interaction strength is of the order of 0.10 meV 
then for temperatures much larger than the energy scale of the 
interaction strength, the system can be essentially regarded 
as a non interacting one. Utilizing this observation, we were 
able to describe the magnetization curves at 1.8 K and 4 K 
(both of which are larger than the expected interaction 
strengths) using just the single ion magnetic field term in 
the Hamiltonian in Eq.~\eqref{eq:Hnn_supp}. 

This was achieved by setting $g_x=0$ and $g_z \approx 2.4$. It was found that while this particular combination of $g_x$ and $g_z$, works very well to describe the high temperature magnetization curves, it was not so accurate in describing the low temperature magnetization at 0.5 K especially at higher values of field strength. (A closer look at the report of Ref.~\onlinecite{CZO.Gao} suggests a $g_z$ value that is changing with magnetic field, an effect not built into our model). Besides this set, we also noticed that the set $g_z\sim 2.35$ and $g_x\sim -0.2324$, which can describe the various magnetization curves reasonably well, has slightly better accuracy while describing the low temperature magnetization curve. This gain in accuracy comes at the cost of a small loss in accuracy when describing the high temperature magnetization curves.

After having found estimates for the $g$ parameters, we minimize the cost function in 
Eq.~\eqref{eq:cost_supp} using a 16 site system by allowing the interaction parameters to vary between different bounds and different initial starting parameters. For this we employed the SLSQP algorithm in the SciPy package. The end results of our investigations yield multiple sets of optimized parameters, they are summarized in Table.~\ref{tab:H_params}. Some representative results for these parameter sets are presented in Fig.~\ref{fig:M_params} and Fig.~\ref{fig:C_params}. Many features of the data of Ref.~\onlinecite{CZO.Gao} are captured correctly, especially in the high temperature and low magnetic field strength regime.  

\begin{figure*}[htpb]
    \centering
    \includegraphics[width=\linewidth]{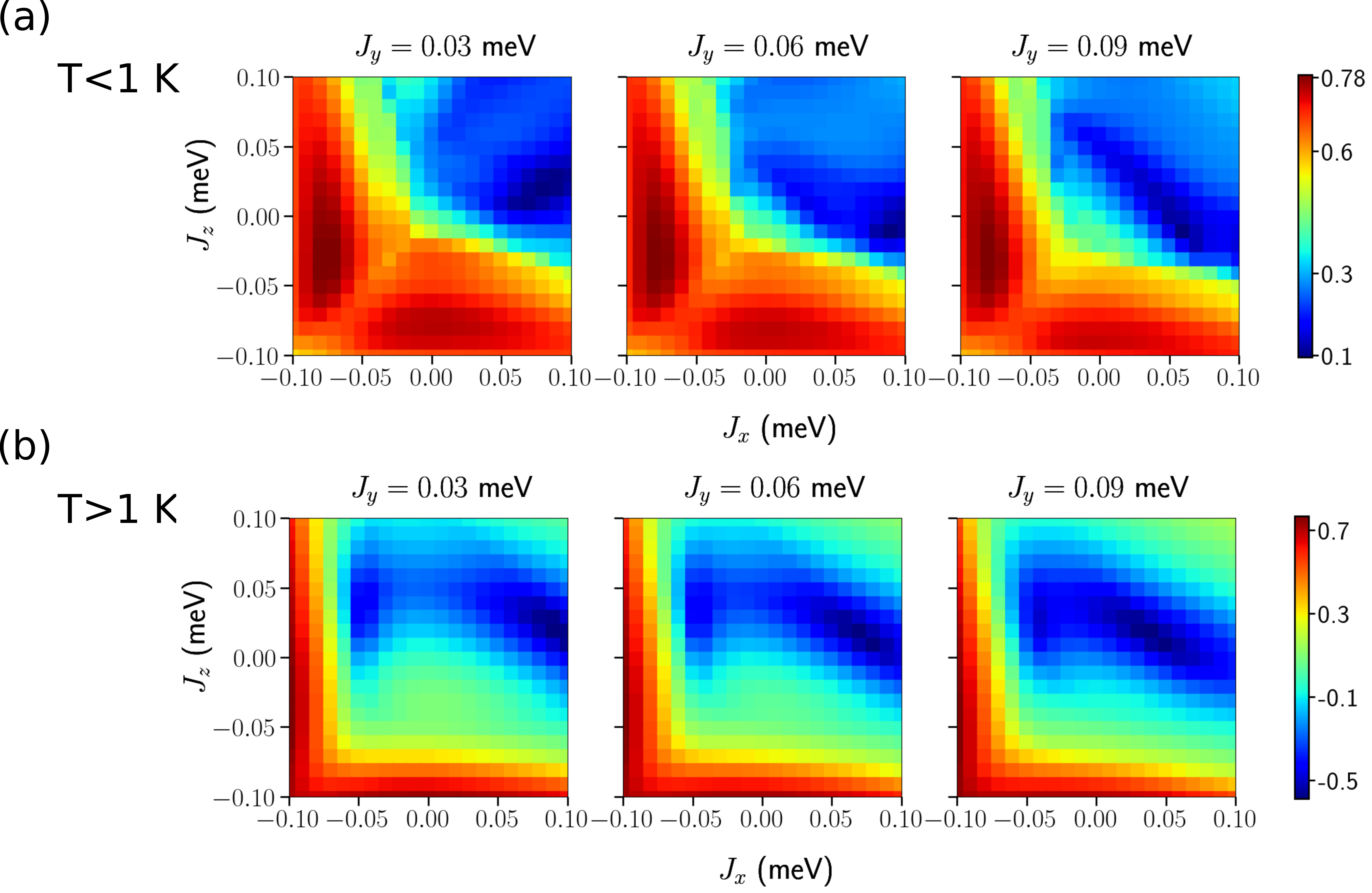}
    \caption{ Two dimensional $J_x-J_z$ cross sections of the cost function for fixed $J_y$-values $0.03$, $0.06$ and $0.09$ meV.  (a) was obtained by using the low temperature part ($T<1$ K) of the specific heat curves in the cost function and (b) was obtained by using their high temperature part ($T>1$ K). The colors represent the log of the cost function. }
    \label{fig:fit1}
\end{figure*}

\begin{figure}[htpb]
\begin{center}
\includegraphics[width=\linewidth]{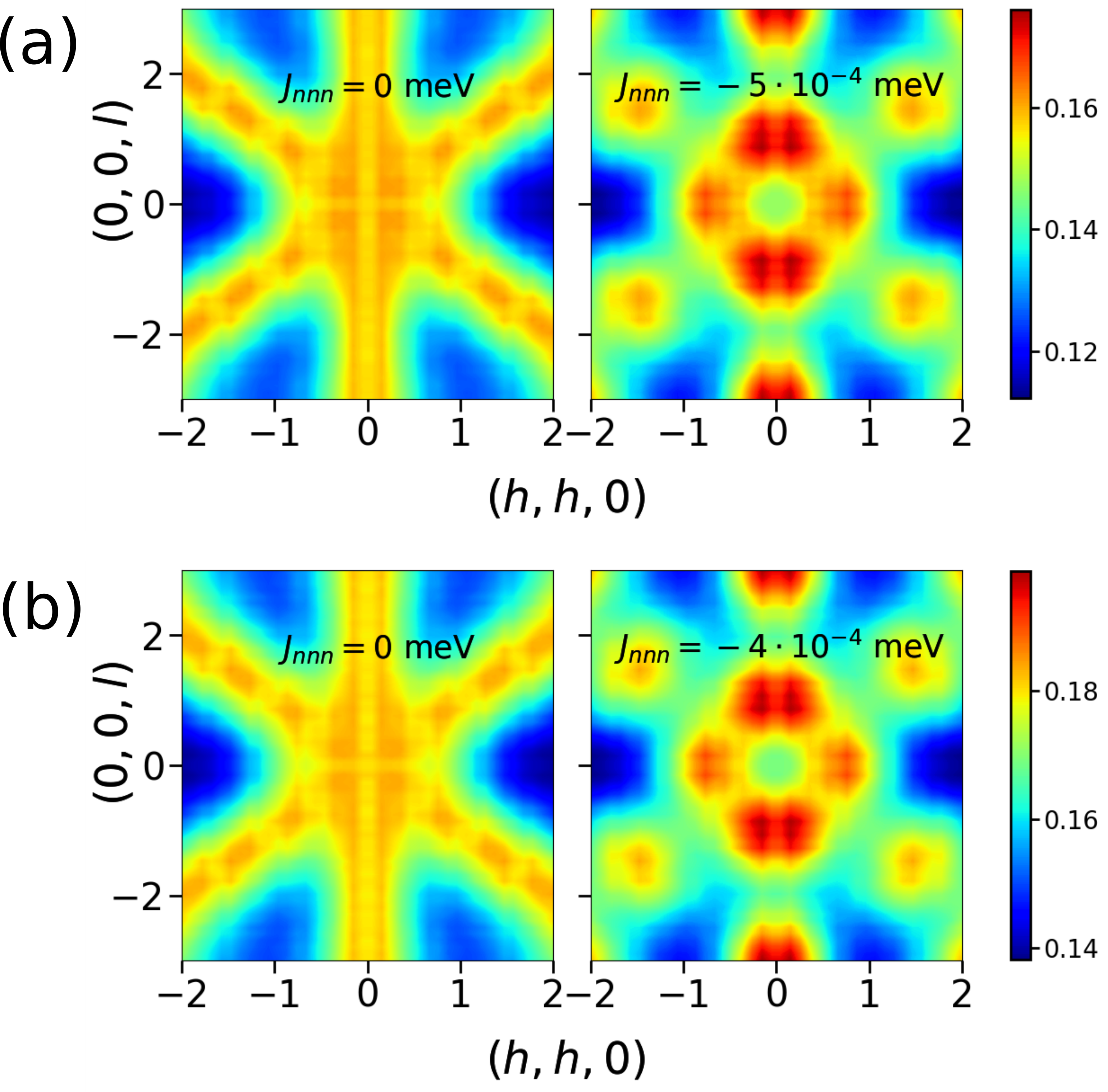}
\caption{Panels (a) and (b) show the (projected) static spin structure factor, obtained by classical Monte Carlo simulation, for parameter set 2 and 4 respectively (see Table~\ref{tab:H_params}). The simulation was performed using 8192 sites and $10^{6}$ Monte Carlo sweeps.}
\end{center}
\label{fig:staticsq} 
\end{figure}

We discuss some more specifics associated with the fitting procedure and the parameter sets.
Parameter set 1 was obtained by fixing $g_x$ and $g_z$ to $0$ and $2.401$ respectively and optimizing the $J_x$, $J_y$, $J_z$ and $J_{xz}$ to fit only the specific heat curves ($\alpha_c=1$ and $\alpha_m=0$). Parameter set 2 was obtained by fixing $g_x$, $g_z$ and $J_{xz}$ to $-0.2324$, $2.35$ and $0$ respectively and then optimizing the $J_x$, $J_y$, $J_z$ to fit the specific heat curves ($\alpha_c=1$ and $\alpha_m=0$). On the other hand, parameter set 3 was obtained by fixing $J_{xz}$ to zero and optimizing everything else to fit both the specific heat and the $T=0.5$ K magnetization curve with field along [111] direction ($\alpha_c=1$ and $\alpha_m=1$). Parameter set 4 was obtained fixing $J_{xz}$ $g_x$ and $g_z$ to zero, zero and 2.4 respectively and optimizing everything else to fit both the specific heat and the $T=0.5$ K magnetization curve with field along [111] direction. It must be noted that all the optimizations were performed using only the high temperature part ($T>1$ K) of the specific heat curves. 

In order to build confidence in our optimized parameters, we also performed a brute force scan in a restricted part of parameter space. We mapped out the cost function for the specific heat ($\alpha_c=1$, $\alpha_m=0$) by varying $J_x$, $J_y$ and $J_z$ from $-0.1$ to $0.1$ meV in steps of $0.01$ meV, while keeping $J_{xz}=0$, $g_x=0$ and $g_z=2.4$. It must be noted that for $g_x=J_{xz}=0$, all properties of the Hamiltonian (including specific heat and magnetization) are invariant to exchanging in $x$ and $y$. Thus the cost function for a set $(J_x,J_y,J_z)$ is identical to that for $(J_y,J_x,J_z)$. 
%(Adding a non zero $g_x$ and/or $J_{xz}$ breaks this symmetry.)

In Fig.~\ref{fig:fit1} we present a few representative cross-sections of the cost-function map  keeping $J_y$ fixed and varying $J_x$. In these maps, the colors correspond to the log of the cost function. $J_y$ was fixed and only either the low temperature ($T<1$ K, Fig.~\ref{fig:fit1}a) or high temperature ($T>1$ K Fig.~\ref{fig:fit1}b) were included in the evaluation of the cost function. Our parameter sets with $g_x=0$ and $g_z=2.4$, which have been obtained using full optimization ($\alpha_c=\alpha_m=1$), lie in the dark blue region of the cost map shown in Fig.~\ref{fig:fit1}b.

We observe that in addition to the solutions obtained by using the optimizer, the cost map suggests existence of other promising sets. This happens because of the $x-y$ symmetry mentioned above. It should be emphasized that this symmetry is only present when $g_x$ and $J_{xz}$ are both zero. In practice, allowing for a non-zero $g_x$ in the optimization selects the $J_y > J_x$ set.
%represent different cross sections where $J_y$ was fixed by and low temperature part ($T<1$ K) and high temperature part ($T>1$ K) of  specific heat curves respectively. 

%Our parameter sets with $g_x=0$ and $g_z=2.4$, which have %been obtained using full optimization ($\alpha=\beta=1$), lie %in the dark blue region of the cost map shown in %Fig.~\ref{fig:fit1}b. 

%We observe that in addition to the solutions obtained by %using the optimizer, the cost map suggests existence of other %promising sets. This happens because of the symmetry, that is %inherently present in the Hamiltonian, by which the value of %the cost function remains unchanged once $J_x$ and $J_y$ are %interchanged. It should be emphasized that this symmetry is %only present when $g_x$ and $J_{xz}$ are both zero.

The optimization procedure also yielded parameter sets which were discarded because their low temperature ($T<1$ K) cost function of specific heat at zero field was large (when compared against the sets mentioned in Table ~\ref{tab:H_params}). Some of these discarded parameter sets have the following interaction strength (in meV): a) $J_x=-0.036$, $J_y=-0.035$ $J_z=0.032$ and $J_{xz}=0$ b) $J_x=-0.043$, $J_y=0.084$, $J_z=0.02$ and $J_{xz}=0$ c)$J_x=0.085$, $J_y=-0.043$, $J_z=0.038$ and $J_{xz}=0$. All our optimizations, using both specific heat and magnetization (both low and high temperature), involving  $J_x$, $J_y$, $J_z$ and $g_z$ (with $J_{xz}=0$ and $g_x=0$)  yielded $g_z\sim 2.2$. Such parameter sets, where $g_z\sim 2.2$  are less reliable as they are unable to explain the magnetization curves accurately at higher temperature (see parameter set 3 in Fig.~\ref{fig:M_params}). 

%The magnetization obtained at $0.5$ K, $1.8$ K, $4.0$ K by using these parameters are shown in Fig.~\ref{fig:mag} and the specific obtained at $4$ T, $8$ T and $14$ T are shown in Fig.~\ref{fig:cv}. 

As mentioned in the main text, $H_{nn}$ alone is insufficient to explain the INS data, at least at the level of a classical treatment of the Hamiltonian. We find that classical Monte Carlo and SCGA calculations of $H_{nn}$ along with a suitably chosen $J_{nnn}$ that enters $H_{nnn}$ (the truncated dipolar interaction discussed earlier in the SM), provides reasonable agreement with the INS data. In Fig.~\ref{fig:staticsq}, we show the static structure factor (to be discussed in the next section) corresponding to what is measured in INS. 
We provide results for Hamiltonian parameter set 2 and 4 both with and without $H_{nnn}$. Since parameter set 1 and 3 are close to parameter set 2, the static structure factor for these sets are also qualitatively similar, and hence not shown. 

\section{Static and Dynamic Structure Factor}

We work in the global $x,y,z$ basis and 
evaluate the equal time (static) expectation value using classical Monte Carlo,
\begin{subequations}
\begin{eqnarray}
S(\qq)&=&\frac{1}{N}\underset{\mu\nu}{\sum}\left(\delta_{\mu\nu}-\frac{q_{\mu}q_{\nu}}{q^2}\right)\langle m^{\mu}(-\qq)m^{\nu}(\qq)\rangle \\
m^{\mu}(\qq)&=&\underset{i}{\sum}e^{-i\qq\cdot\vec{r}_i}\underset{\lambda}{\sum}g_i^{\mu\lambda}S_i^{\lambda} 
\end{eqnarray} 
\end{subequations}
where $N$ is the number of sites. We employ the usual Metropolis algorithm with continuous conical moves to sample spin configurations. We perform $N$ single spin moves (collectively referred to as a sweep) before making a measurement.  

For dynamical properties, we use the molecular dynamics (MD) procedure. In this method, $N_{IC}$ equilibrium configurations were first drawn from the thermal ensemble at $T=0.06$ K using Monte Carlo, then each of these initial configurations (IC) were evolved in time by using the Landau-Lifshitz equation, 
\\
\begin{equation}
\frac{d {\bf{S_i}}} {dt}= {\bf{S_i}} \times {\bf h_{eff,i}}
\label{eq:LL_supp}
\end{equation} 
\\
where ${\bf{h_{eff,i}}}$ is the effective magnetic field (local exchange field) experienced by a spin at site $i$, due to the interactions with all other spins it is coupled to. The time evolution of spins given by Eq.~\eqref{eq:LL_supp} is performed numerically using the fourth order Runge-Kutta method~\cite{Keren, Conlon_Chalker,ncnf.zhang}. A sufficiently small time step was chosen to ensure that the energy was (roughly) constant with time. The evolution was done for a total time of $T_{s}=500$ meV$^{-1}$. 

To compare with what is measured in the INS experiment, we first measure the appropriately projected dynamical structure factor for each IC (which we refer to as $S^{IC}(\qq,\omega)$) 
and average it over $N_{IC}$ configurations:
\begin{widetext}
\begin{subequations}
\begin{eqnarray}
S^{IC}(\qq,\omega)&=& \frac{T_s}{\pi N} \underset{\mu\nu}{\sum}\left(\delta_{\mu\nu}-\frac{q_{\mu}q_{\nu}}{q^2}\right) \langle m^{\mu}(-\qq,-\omega)m^{\nu}(\qq,\omega)\rangle \\
m^{\mu}(\qq,\omega)&=&\frac{1}{T_s}\int_0^{T_s} dt\; e^{i\omega t} \underset{i}{\sum}e^{-i\qq\cdot\rr_i}\underset{\lambda}{\sum}g_i^{\mu\lambda}S_i^{\lambda} 
\end{eqnarray}
\end{subequations}
\end{widetext}
We then use the quantum-classical correspondence discussed in Ref.~\onlinecite{ncnf.zhang} to obtain $S_{quantum}(\qq,\omega) \equiv \beta \omega S(\qq,\omega)$

\section{Importance of $J_y$}
In most parameter sets, in particular set no. 2 (see Table~\ref{tab:H_params}) which was used for the calculations in the main text, we observe that $J_y$ is the dominant interaction term. To assess its importance, we evaluate the specific heat and magnetization by using this parameter set with $J_y$ set to zero. The final result obtained has been compared with experiment, see Fig.~\ref{figap:0Jy}. We observe that the specific heat in zero applied field, obtained from parameter set 2 with $J_y$ set to zero, is unable to describe the experiment at higher values of temperature. We also observe that the [111] magnetization at 0.5 K obtained with $J_y=0$ is less accurate at lower magnetic field strengths.
\begin{figure}[htpb]
            \includegraphics[width=\linewidth]{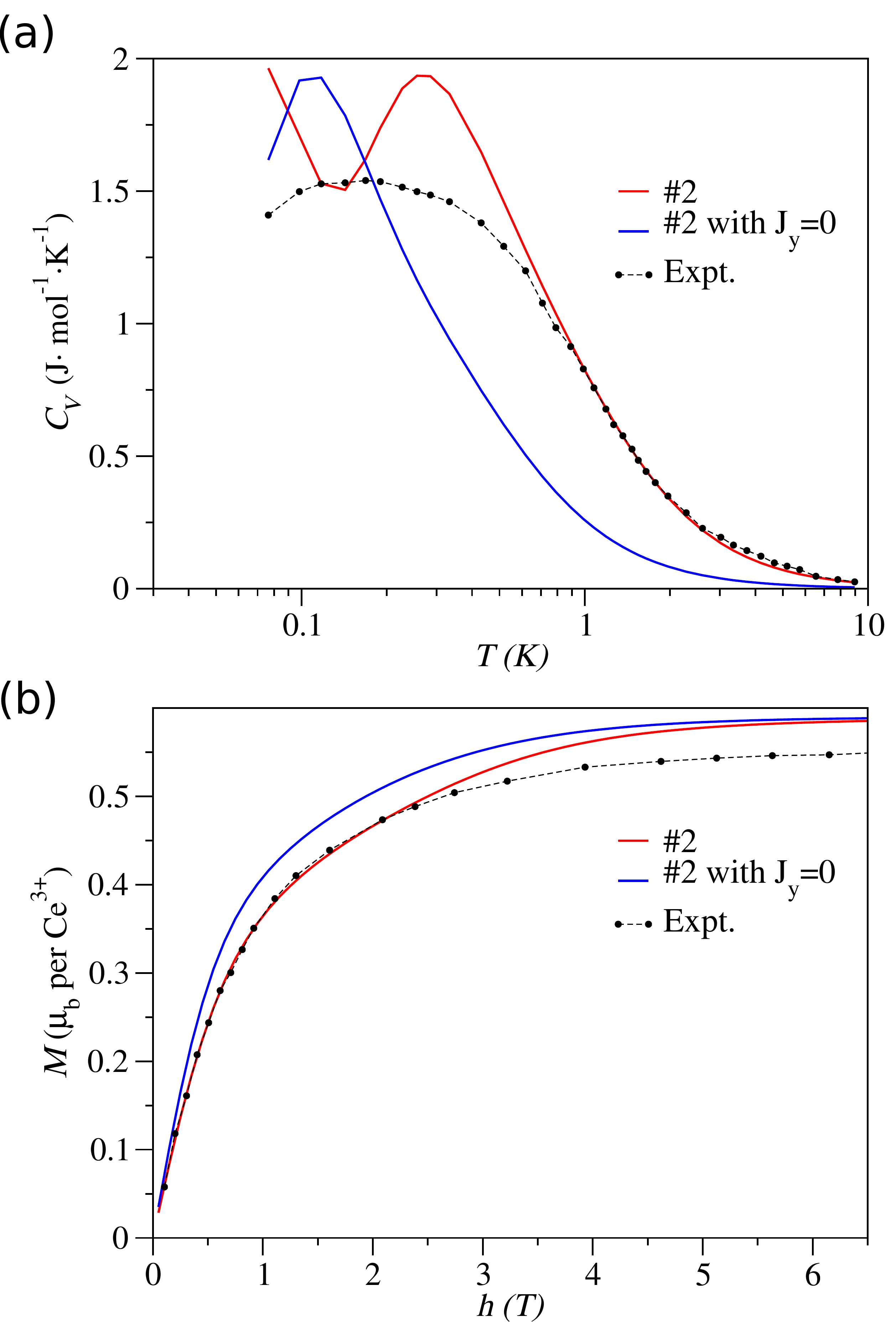}
\caption{\label{figap:0Jy}(a) Comparison of the specific heat vs. temperature profile at zero field obtained by using parameter set 2 (red), parameter set 2 with $J_y$ set to zero (blue) and the experiment (black solid circles connected by dashed lines). (b) Magnetization vs. magnetic field along the [111] direction at 0.5 K.} 
%obtained by using parameter set 2 (red), parameter set 2 with %$J_y$ set to zero (blue) and the experiment respectively %(black solid circles connected by dashed lines).} 
\end{figure}

\section{Single ion physics}
For a single electron in the presence of uniform magnetic field, the leading order interaction (in orders of field strength $h$), is the Zeeman term given by,
\begin{eqnarray}
H_{Z}= - {\bf  h \cdot(L +2 S)}.
\label{zman}
\end{eqnarray}
When the ground state is well isolated from the excited states, the low energy physics of the ion can be captured by expressing the above Hamiltonian in the subspace spanned by the lowest lying multiplet. Here, we discuss two cases, namely (a) the ground state is a doublet  which is given  $J=5/2$ and $m_J=\pm 3/2$ and (b) the ground state is a doublet which is a particular linear combination of the $J=5/2$ and $J= 7/2$. 
We refer to the local quantization axis as $z$ which coincides with the local [111] axis.

\subsection{Ground state doublet is $|J=5/2,\;m_J=\pm 3/2 \rangle$}

We first consider the case where the ground state doublet is 
\begin{eqnarray}
|\pm\rangle=|5/2,\pm 3/2\rangle.
\label{groundstate1}
\end{eqnarray}
We treat the term mentioned in Eq.~\eqref{zman} as a perturbation and find its matrix elements in the $|\pm \rangle$ subspace. To do so, we make use of a result that follows from the Wigner-Eckart\cite{sakurai1985modern} theorem, which states that the matrix element of any vector operator in the eigenstates of $\bf{J}^2$ and $J_z$ with a given $J$ are proportional to the matrix element of ${\bf J}$ itself.
\begin{eqnarray}
\langle J,J_z|{\bf L +2 S} |J,J_z'\rangle=g(JLS)\langle J,J_z|{\bf J}|J,J_z'\rangle.
\label{WET}
\end{eqnarray}
where,
\begin{eqnarray}
g(JLS)=\frac{3}{2}+\frac{1}{2}\left[\frac{S(S+1)-L(L+1)}{J(J+1)}\right].
\end{eqnarray}

Clearly, $\langle\pm|{\bf L +2 S}|\mp\rangle=0$ and only the diagonal elements contribute. For the diagonal elements we note that in the matrix element $\langle +|{\bf L + 2 S}|+\rangle$ only the $z$-component of the vector contributes. This means 
\begin{eqnarray}
{\bf h} \cdot\langle +|{\bf L + 2 S}|+\rangle&=& h_z\langle +|L_z+2S_z|+\rangle \nonumber\\ 
&=& h_zg(JLS)\langle +|J_z|+\rangle.
\end{eqnarray}
The matrix element ${\bf h} \cdot\langle -|{\bf L + 2 S}|-\rangle$ can be evaluated in a similar way, and thus the Zeeman term in the $|\pm\rangle$ basis takes the form, % of a 2 by 2 traceless-diagonal matrix given by
\begin{eqnarray}
H_Z= - h_z g_z s_z.
\end{eqnarray}
where $s_z=\frac{\sigma_z}{2}$ is a an effective spin-1/2 operator and $\sigma_z$ is the usual $2 \times 2$ Pauli matrix.

\subsection{Ground state doublet is a linear combination of $J=5/2$ and $J=7/2$}
For the isostructural compound Ce$_2$Sn$_2$O$_7$ Ref.~\onlinecite{natphys.Sibille2018} has determined that the ground state doublet, after including the $J=7/2$ manifold, for a single Cerium ion in the presence of spin-orbit coupling and crystal field is,
\begin{eqnarray}
\label{eq:sibille}
|\pm\rangle&=&0.87|^2F_{5/2},\pm 3/2\rangle\pm 0.46|^2F_{5/2},\mp 3/2\rangle \nonumber \\ 
&& \mp 0.15|^2F_{7/2},\pm 3/2\rangle -0.01|^2F_{7/2},\mp 3/2\rangle.
\end{eqnarray}
 We note that Ref.~\onlinecite{CZO.Gao} has not reported any such mixing for the case of 
 Ce$_2$Zr$_2$O$_7$.
 %This is contrary to the case of Ce$_2$Zr$_2$O$_7$ where the ground state of a single Cerium ion was determined by using only the $J=5/2$ manifold. 
 We allow for such a possibility for Ce$_2$Zr$_2$O$_7$, and leave the precise determination of the wavefunction coefficients to future experiments. Instead we will use the numbers appearing in Eq.~\eqref{eq:sibille} %In this section we will assume that the ground state of Cerium ion in Ce$_2$Zr$_2$O$_7$ has the same form as in~\eqref{sibille} and proceed 
 simply to motivate the form of the Zeeman term in the subspace of this doublet. The exercise will illustrate the origins of non-zero $g_x$ and $g_z$.
 
Clearly, Eq.~\eqref{WET} can no longer be used to evaluate the matrix element of the type $\langle^2F_{5/2},\pm 3/2|{\bf L + 2 S}|^2F_{7/2},\pm 3/2\rangle$ as the $J$-values on the left and right are different.
To evaluate these matrix elements we first define,
\begin{subequations}
\begin{eqnarray}
|1\rangle& \equiv &|^2F_{5/2},3/2\rangle, \\
|2\rangle& \equiv &|^2F_{5/2},-3/2\rangle, \\
|3\rangle& \equiv &|^2F_{7/2},3/2\rangle, \\
|4\rangle& \equiv &|^2F_{7/2},-3/2\rangle. 
\end{eqnarray}
\end{subequations}
It is easy to observe that
\begin{eqnarray}
\langle i|L_{\alpha}+2S_{\alpha}|j\rangle=0,~~i,j~\epsilon~(1,2,3,4),~~\alpha~\epsilon~(x,y).
\end{eqnarray}
So that the only terms needed to be evaluated are $\langle i|L_z+2S_z|j\rangle$. It can also be shown that
$\langle 1|L_z+2S_z|2\rangle = \langle 1|L_z+2S_z|4\rangle
= \langle 2|L_z+2S_z|3\rangle = \langle 3|L_z+2S_z|4\rangle = 0.$ 
%\end{eqnarray}
%\end{subequations}
%\begin{subequations}
%\begin{eqnarray}
%\langle 1|L_z+2S_z|2\rangle&=&0, \\
%\langle 1|L_z+2S_z|4\rangle&=&0, \\
%\langle 2|L_z+2S_z|3\rangle&=&0, \\
%\langle 3|L_z+2S_z|4\rangle&=&0. 
%\end{eqnarray}
%\end{subequations}
The remaining combination can be evaluated by using 8 different Clebsch-Gordon coefficients which are determined by deriving the following relations,
\begin{widetext}
\begin{subequations}
\begin{eqnarray}
|J=\frac{5}{2},J_z=+\frac{3}{2}\rangle&=&+\sqrt{\frac{5}{7}}|L=3,L_z=+2;S=\frac{1}{2},S_z=-\frac{1}{2}\rangle-\sqrt{\frac{2}{7}}|L=3,L_z=+1;S=\frac{1}{2},S_z=+\frac{1}{2}\rangle, \\
|J=\frac{5}{2},J_z=-\frac{3}{2}\rangle&=&-\sqrt{\frac{5}{7}}|L=3,L_z=-2;S=\frac{1}{2},S_z=+\frac{1}{2}\rangle+\sqrt{\frac{2}{7}}|L=3,L_z=-1;S=\frac{1}{2},S_z=-\frac{1}{2}\rangle, \\ 
|J=\frac{7}{2},J_z=+\frac{3}{2}\rangle&=&+\sqrt{\frac{2}{7}}|L=3,L_z=+2;S=\frac{1}{2},S_z=-\frac{1}{2}\rangle+\sqrt{\frac{5}{7}}|L=3,L_z=+1;S=\frac{1}{2},S_z=+\frac{1}{2}\rangle, \\
|J=\frac{7}{2},J_z=-\frac{3}{2}\rangle&=&+\sqrt{\frac{2}{7}}|L=3,L_z=-2;S=\frac{1}{2},S_z=+\frac{1}{2}\rangle+\sqrt{\frac{5}{7}}|L=3,L_z=-1;S=\frac{1}{2},S_z=-\frac{1}{2}\rangle. 
\end{eqnarray}
\end{subequations}
\end{widetext}
From the above equations we obtain,
\begin{subequations}
\begin{eqnarray}
\langle 1|L_z+2S_z|1\rangle&=& - \langle 2|L_z+2S_z|2\rangle 
=  \frac{9}{7}, \\
\langle 3|L_z+2S_z|3\rangle&=& -\langle 4|L_z+2S_z|4\rangle= \frac{12}{7}, \\
\langle 1|L_z+2S_z|3\rangle&=&\langle
3|L_z+2S_z|1\rangle = -\frac{\sqrt{10}}{7},\\
\langle 2|L_z+2S_z|4\rangle&=&\langle 4|L_z+2S_z|2\rangle = -\frac{\sqrt{10}}{7}.
\end{eqnarray}
\end{subequations}

Using the above equations we find that,
\begin{subequations}
\begin{eqnarray}
\langle +|L_z+2S_z|+\rangle&=&0.8615, \\
\langle -|L_z+2S_z|-\rangle&=&-0.8615, \\
\langle +|L_z+2S_z|-\rangle&=&\langle -|L_z+2S_z|+\rangle=-1.0784.  
\end{eqnarray}
\end{subequations}
It is interesting to note that $\langle +|L_z+2S_z|-\rangle$ is not zero (unlike the case of pure $J=5/2$ ground state doublet) and that the matrix elements of ${\bf L+2S}$ only depend on the $z$-component. Using the above equations it immediately follows that the Zeeman term reduces to the form,
%We note that the matrix element of ${\bf L +2 S}$ only depends on its $z$-component
%\begin{eqnarray}
%{\bf h\cdot\langle \pm|L +2 S |\pm\rangle}=h_z\langle \pm|L_z+2S_z|\pm\rangle
%\end{eqnarray}

\begin{eqnarray}
H_Z = -h_z(g_x s_x+g_z s_z).
\end{eqnarray}
when expressed in the subspace of the ground state doublet.

\section{The Hamiltonian as a $\pi$-flux octupolar spin liquid}
 
In this section we briefly explain how the parameters 
of Table~\ref{tab:H_params}. 
place the model in the phase of $\pi-$flux octupolar quantum spin ice, and the meaning of  ``$\pi-$flux''. 

For $J_{xz}=0$ and no external magnetic field, the nn Hamiltonian in the local basis is
written as 
\begin{equation}
\mathcal{H} = \sum_{\langle ij \rangle} \left[ J_y s_i^z s_j^z 
+ J_x s_i^x s_j^x 
+ J_z s_i^z s_j^z   \right ] .
\end{equation}
First, we notice that 
$J_y \gg J_x,\ J_z$ in most of the fitted parameters (here we take parameters $J_y=0.08$ meV, $J_x=0.05$ meV, $J_z=0.02$ meV). 
Hence we treat the term $J_y s_i^z s_j^z $ as the dominating one. 
It enforces the ``2-in-2-out'' ice rule
on each tetrahedron for the local $s^y$
components.
Since $s^y$ is of octupolar nature,
the large $J_y$ places the system in 
the octupolar ice phase. 
 
The other terms $J_x s_i^x s_j^x 
+ J_z s_i^z s_j^z$ introduce quantum dynamics to the octupolar ice states.
We can rewrite them in terms of the raising
and lowering operators 
of $s^y$
\begin{equation}
    s^{y+}=s^z + i s^x,
    \quad
    s^{y-}=s^z - i s^x.
\end{equation}
The Hamiltonian then becomes
\begin{equation}
\begin{split}
\mathcal{H} = 
\sum_{\langle ij \rangle} 
&\bigg[ J_y s_i^y s_j^y + 
 \frac{J_z + J_x}{4} (s^{y+}_i s^{y-}_j+ s^{y+}_j s^{y-}_i) \\
&+
 \frac{J_z - J_x}{4}  (s^{y+}_i s^{y+}_j + s^{y-}_j s^{y-}_i) \bigg ] .
 \end{split}
\end{equation}
where
\begin{equation}
    0 <   \frac{J_z + J_x}{4} \ll J_y,
    \quad
     \frac{J_x - J_z}{4}  \ll   \frac{J_z + J_x}{4} .
\end{equation} 

If we ignore the term with the smallest coefficient  $\frac{J_x - J_z}{4}$,
the rest of the Hamiltonian 
becomes identical to
that of regular quantum spin ice model,
and has been studied in detail.
The term $ \frac{J_z + J_x}{4} (s^{y+}_i s^{y-}_j+ s^{y+}_j s^{y-}_i) $
flips two neighbouring spins.
When restricted to the ``2-in-2-out''
ice state Hilbert space,
it perturbatively generates the loop exchange term.
The lowest order one is defined on  hexagons of the pyrochlore lattice as
\begin{equation}
H_\text{flux} = J_\text{ring} \sum_{\hexagon} \cos(\nabla \times a)_{\hexagon,}
\end{equation}
where $J_\text{ring} \sim (J_x + J_z)^3/(64 J_y^2)$ is positive. 
 
The quantum dynamical terms are believed to lead the system into a quantum spin ice phase, as shown in various studies.
Furthermore, 
the positive $J_\text{ring}$ favors the ground state to satisfy $\cos(\nabla \times  \bm{A})   =  -1$ on each hexagon,
which is refereed to as  $\pi-$flux state.
This ground state is qualitatively different from the phase of $0-$flux state favoured by $\cos(\nabla \times  \bm{A})   =   1$.
In the $\pi-$flux state, the mean-field ansats of gauge fields 
is not trivially $\bf{A}=0$ but  a more complex pattern\cite{Lee2012PhysRevB},
and leads the system into a different quantum spin ice phase than that of $\bf{A}=0$ background.
More detailed study of the  $\pi-$flux  phase can be found in Ref.~\onlinecite{Lee2012PhysRevB,Benton2018PhysRevLett,GangChen2017PhysRevB,Patri-octupolarQSI}. 

The $\pi-$flux quantum spin ice is expected to be more stable than the $0-$flux quantum spin ice. Crudely speaking, this is due to the positive $J_{x,y,z}$ create more frustration than just one positive $J_y$, and stabilizes the phase to a large region of the parameter space.
So turning on the small $\frac{J_z - J_x}{4}  (s^{y+}_i s^{y+}_j + s^{y-}_j s^{y-}_i)$ term does not qualitatively affect the physics, as shown in the phase diagram in Ref.~\onlinecite{Patri-octupolarQSI}.

\end{document}